%% file: main.tex
\documentclass[10pt,conference]{IEEEtran}
\usepackage{cite}
\usepackage{amsmath,amssymb,amsfonts}
\usepackage{algorithmic}
\usepackage{graphicx}
\usepackage{textcomp}
\usepackage{xcolor}
\usepackage[hyphens]{url}
\usepackage[T1]{fontenc}

\usepackage{enumitem}
\usepackage{flushend}
\usepackage{balance}
\usepackage{booktabs}
\usepackage{epsfig}
\usepackage{graphicx}
\usepackage{fancyhdr}
\usepackage{amsmath}
\usepackage{amssymb}
\usepackage{lipsum}
\usepackage{subfig}
\usepackage{color}

\usepackage[normalem]{ulem}
\usepackage{bigstrut}
\usepackage{multirow}
\usepackage{siunitx}
\usepackage{ulem}
\usepackage{listings}
\usepackage{eso-pic}
\usepackage{tikz}
\usepackage{xspace}
\usepackage{fancyhdr}
\usepackage{mathtools}
\usepackage{soul}
\usepackage[leftcaption]{sidecap}

\usepackage{fancyhdr}
\usepackage{kantlipsum}
\usepackage{eso-pic}

\usepackage[bookmarks=true,breaklinks=true,letterpaper=true,colorlinks,citecolor=blue,linkcolor=blue,urlcolor=blue]{hyperref}
\def\BibTeX{{\rm B\kern-.05em{\sc i\kern-.025em b}\kern-.08em
    T\kern-.1667em\lower.7ex\hbox{E}\kern-.125emX}}

\title{Thales: Formulating and Estimating Architectural Vulnerability Factors for DNN Accelerators}

\makeatletter
\newcommand{\linebreakand}{%
  \end{@IEEEauthorhalign}
  \hfill\mbox{}\par
  \mbox{}\hfill\begin{@IEEEauthorhalign}
}
\makeatother
\author{\IEEEauthorblockN{Abhishek Tyagi}
\IEEEauthorblockA{
\textit{University of Rochester}\\
atyagi2@ur.rochester.edu}
\and
\IEEEauthorblockN{Yiming Gan}
\IEEEauthorblockA{\textit{University of Rochester} \\
ygan10@ur.rochester.edu}
\and
\IEEEauthorblockN{Shaoshan Liu}
\IEEEauthorblockA{\textit{PerceptIn} \\
shaoshan.liu@perceptin.io}
\and
\IEEEauthorblockN{Bo Yu}
\IEEEauthorblockA{
\textit{PerceptIn}\\
bo.yu@perceptin.io}
\linebreakand
\IEEEauthorblockN{Paul Whatmough}
\IEEEauthorblockA{
\textit{Arm Research}\\
paul.whatmough@arm.com}
\and
\IEEEauthorblockN{Yuhao Zhu}
\IEEEauthorblockA{
\textit{University of Rochester}\\
yzhu@rochester.edu}
}
\input{macro}
\graphicspath{{figs/}}
\begin{document}
\AddToShipoutPictureBG*{%
  \AtPageUpperLeft{%
    \setlength\unitlength{1in}%
    \hspace*{\dimexpr0.5\paperwidth\relax}
    \makebox(0,-0.75)[c]{\small This article appears in IEEE International Symposium on High-Performance Computer Architecture (HPCA), 2023}%
}}

\maketitle
\thispagestyle{plain}
\pagestyle{plain}


\input{abst}
\input{intro}

\input{moti}

\input{robust}
\input{sample}
\input{validation}

\input{eval}
\input{compare}
\input{nas}

\input{relate}
\input{concl}

\bibliographystyle{IEEEtranS}
\bibliography{refs}

\end{document}

%% file: macro.tex
\ifx\figurename\undefined \def\figurename{Figure}\fi
\renewcommand{\figurename}{Fig.}
\renewcommand{\paragraph}[1]{\textbf{#1} }

\newcommand{\Sect}[1]{Sec.~\ref{#1}}
\newcommand{\Fig}[1]{Fig.~\ref{#1}}
\newcommand{\Tbl}[1]{Tbl.~\ref{#1}}
\newcommand{\Equ}[1]{Equ.~\ref{#1}}

\newcommand{\proj}{\textsc{Thales}\xspace}

\newcommand{\mode}[1]{\underline{\textsc{#1}}\xspace}

\newcommand{\no}[1]{#1}
\renewcommand{\no}[1]{}
\newcommand{\RNum}[1]{\uppercase\expandafter{\romannumeral #1\relax}}


%% file: abst.tex
\begin{abstract}

As Deep Neural Networks (DNNs) are increasingly deployed in safety-critical and privacy-sensitive applications such as autonomous driving and biometric authentication, it is critical to understand the fault-tolerance nature of DNNs.
Prior work primarily focuses on metrics such as Failures In Time (FIT) rate and the Silent Data Corruption (SDC) rate, which quantify \textit{how often} a device fails. Instead, this paper focuses on quantifying the DNN accuracy \textit{given} that a transient error has occurred, which tells us how well a network behaves when a transient error occurs. We call this metric Resiliency Accuracy (RA).




We show that existing RA formulation is fundamentally inaccurate, because it incorrectly assumes that software variables (model weights/activations) have equal faulty probability under hardware transient faults. We present an algorithm that captures the faulty probabilities of DNN variables under transient faults and, thus, provides correct RA estimations validated by hardware. To accelerate RA estimation, we reformulate RA calculation as a Monte Carlo integration problem, and solve it using importance sampling driven by DNN-specific heuristics.

Using our lightweight RA estimation method, we show that transient faults lead to far greater accuracy degradation than what today's DNN resiliency tools estimate. We show how our RA estimation tool can help design more resilient DNNs by integrating it with a Network Architecture Search framework.

\end{abstract}

%% file: intro.tex
\section{Introduction}
\label{sec:intro}

Deep neural networks (DNNs) have become integral components in many safety-critical and/or privacy-sensitive application domains\no{ such as autonomous machines~\cite{rao2018deep,maqueda2018event,ni2020survey,gupta2021deep}, biometric authentication~\cite{everson2018biometricnet,al2018multi,sundararajan2018deep}, and medical diagnosis~\cite{bakator2018deep,shen2017deep,liu2019deep,azad2021medical}}. Thus, the fault tolerance of a deep learning system has become crucial. 
Among many sources of vulnerability, a major concern is transient errors, which are radiation-induced accidental bit flips~\cite{seifert2006radiation,seifert2010radiation,srinivasan1994accurate,papadimitriou2021demystifying}. While transient bit flips can occur anywhere on a chip~\cite{naseer2007critical,jahinuzzaman2009soft,degalahal2003analyzing,lantz1996soft,de2015evaluation}, they are particularly detrimental to flip-flops (FFs); other memory structures (SRAMs and DRAMs) are usually protected by error correction codes~\cite{slayman2005cache,chen1984error}.

Prior work primarily focuses on metrics such as Failures In Time (FIT) rate and the Silent Data Corruption (SDC) rate, which quantify \textit{how often} a device fails. Instead, this paper focuses on quantifying the DNN accuracy \textit{given} that a transient error has occurred, which tells us how well a network behaves when a transient error occurs. We call this metric Resiliency Accuracy (RA), the counterpart of the network's fault-free, Standard Accuracy (SA).

The RA metric can be seen as the Architectural Vulnerability Factor (AVF)~\mbox{\cite{mukherjee2003systematic, leveugle2009statistical}} of a DNN accelerator.
As traditionally defined (for CPU and GPU architectures), the AVF of a fault site (i.e., a single bit in a single cycle) is either 0 (the value of that bit in that cycle does not affect the correctness of the program) or 1 (the value of that bit in that cycle does affect the correctness of the program); this is averaged across all bits and cycles to get an average AVF for an application.
In contrast, the AVF of a fault site for a DNN accelerator is not binary, because DNN's accuracy (over a test set) is a percentage. Thus, a DNN accelerator's AVF is a spectrum between 0 and 1.

Prior work conducted real beam experiments on GPUs~\mbox{\cite{dos2018analyzing}} and TPUs~\mbox{\cite{rech2022reliability}} to demonstrate the DNN accuracy loss under transient faults. The focus of our work is to show an approach that estimates RA using fault injections. Existing DNN fault injection tools such as PyTorchFI~\mbox{\cite{mahmoud2020pytorchfi}}, Ares~\mbox{\cite{reagen2018ares}}, TensorFI~\mbox{\cite{chen2020tensorfi}} inject faults into DNN variables (i.e., weights and activations) and average the resulting inference accuracies. This approach resembles the classic Software Vulnerability Factor (SVF) analysis~\mbox{\cite{sridharan2009eliminating, madeira2000emulation, wei2014quantifying}}, and does not accurately reflect the AVF because it fundamentally assumes that DNN variables are equally faulty.
We show that the faulty probability varies by orders of magnitude across DNN variables (\mbox{\Sect{sec:motivation}}), leading to the mismatch between AVF and SVF in DNNs; our results corroborate and complement the CPU-centric results in Papadimitriou and Gizopoulos~\mbox{\cite{papadimitriou2021demystifying}}.

Following how AVF is defined for classic architectures, we define RA as an \mbox{\textit{expected value}} of the DNN's inference accuracy, which weighs the inference accuracy by how probable a software variable receives a transient fault at each cycle. Crucially, each software variable has a potentially different faulty probability.
In resiliency parlance, our RA metric is AVF modulated by the impact of the bit flip on DNN accuracy.


We then describe how the new RA metric can be analytically derived from simple statistics of a DNN and the accelerator. This is achieved by leveraging the regular data reuse pattern in DNNs to map a software variable to hardware FFs over time (\Sect{sec:ptf}).
We show that our analytical modeling method correlates well with results obtained from large-scale (over 2.6 billion) RTL fault simulations (\Sect{sec:val}).


Accurately calculating RA using the new formulation is time-consuming: it requires exhaustively enumerating all the bits in all the model weights/activations, each of which requires inferencing over an entire test set. We resort to sampling. Critically, the faulty probabilities in software vary significantly; uniform sampling, a common strategy used in prior work, would lead to significant estimation variance or require an excessive amount of samples to converge.

We propose a sampling strategy that estimates the true RA with several orders of fewer samples (\Sect{sec:sampling}). The key is to formulate RA estimation as a Monte Carlo integration problem~\cite{press2007numerical}. Taking this perspective, we propose to use importance sampling~\cite{kloek1978bayesian} to reduce estimation variance and accelerate convergence. We show that a perfect estimator based on important sampling is impossible, and propose heuristics that 
leverage DNN-specific characteristics to approximate importance sampling while
retaining the main benefit.
We show that our importance sampling method significantly accelerates the convergence rate of RA estimation (\Sect{sec:ra}).

We apply our RA formulation and estimation method for a set of case studies. For instance, we show that transient errors lead to much a higher DNN accuracy degradation compared to what today's RA formulation estimates; selectively hardening control FFs improves RA the most (\Sect{sec:compare}).
We also show how our RA formulation helps design DNNs that are more resilient to transient faults by integrating our RA estimation into a Network Architecture Search (NAS) framework (\Sect{sec:nas}).

In summary, our paper makes the following contributions:

\begin{itemize}
    \item We propose a new formulation of Resiliency Accuracy (RA) to quantify a DNN's accuracy under transient faults.
    \item We propose a method to estimate RA in software. The method formulates RA estimation as a Monte Carlo integration problem and uses DNN-specific importance sampling to solve the integration problem.
    \item We apply our RA formulation to common DNNs and show that their accuracy loss is far greater than what is estimated by today's RA metric.
    \item We demonstrate how our RA formulation can help design more resilient DNNs when coupled with NAS.
\end{itemize}

%% file: moti.tex
\section{Background and Motivation}
\label{sec:motivation}

We first define the scope of transient faults this paper focuses on (\Sect{sec:motivation:scope}). We then discuss the intuition behind needing a resiliency accuracy metric (\Sect{sec:motivation:metric}), followed by describing the common formulation of the metric in existing tools  (\Sect{sec:motivation:sw}). We quantitatively analyze why the existing formulation is fundamentally flawed (\Sect{sec:motivation:errors}).


\subsection{Scope and Assumptions}
\label{sec:motivation:scope}

Consistent with recent prior work such as FIdelity~\mbox{\cite{he2020fidelity}} and Mahmoud et al.~\mbox{\cite{mahmoud2021optimizing}}, we consider only single-bit errors in FFs, which are ``\textit{the most prominent abstraction for transient errors including soft errors and voltage variations.}''~\cite{he2020fidelity}. We assume that other major memory structures (e.g., DRAM, SRAM) are protected by ECC and/or parity bits~\cite{mahmoud2021optimizing}. Note that FF faults capture not only transients faults in FFs but also faults taken place in logic computation.

As with Mahmoud et al.~\cite{mahmoud2021optimizing}, we also assume that different bit positions in an FF have the same raw FIT rate\footnote{Section V-A: ``\textit{each error injection is performed on a single random bit of a random neuron for a random image.}''}. In real designs, some FF bits (e.g., exponent in float) could be hardened to resist transient errors~\cite{kobayashi2014low}.
Note, however, that we do \textit{not} assume that all the FFs have the same fault probability. In fact, each FF (or each category of FFs) is characterized by its raw FIT rate, which will participate in our new metric, as we will discuss in \Sect{sec:robust:def}.

\subsection{Why a New Metric?}
\label{sec:motivation:metric}

Failures In Time (FIT) rate and the Silent Data Corruption (SDC) rate of a device (accelerator) are common metrics used in the fault tolerance literature for quantifying the impact of transient errors~\mbox{\cite{feng2010shoestring, hari2012low, chen2019binfi}}. These metrics quantify \mbox{\textit{how often}} a device fails. In the context of DNN accelerators, an SDC is an inference mis-prediction~\mbox{\cite{chen2020tensorfi}}, and the FIT rate calculation considers not only mis-predictions but also system anomalies such as a crash~\mbox{\cite{he2020fidelity}}.

This paper focuses on an orthogonal metric, which quantifies the accuracy of a DNN \textit{given} that a transient error has occurred. We call this metric Resiliency Accuracy (RA), the counterpart of the network's fault-free inference accuracy, which we call Standard Accuracy (SA).
Intuitively, RA tells us how well/badly a network behaves \textit{when} a transient error occurs. Knowing RA lets us develop DNN algorithms and accelerators that can still perform reasonably well even when the execution is ``corrupted'' by soft errors.

\begin{figure*}[t]
\centering
\subfloat[\small{LeNet.}]
{
  \includegraphics[trim=0 0 0 0, clip, width=0.9\columnwidth]{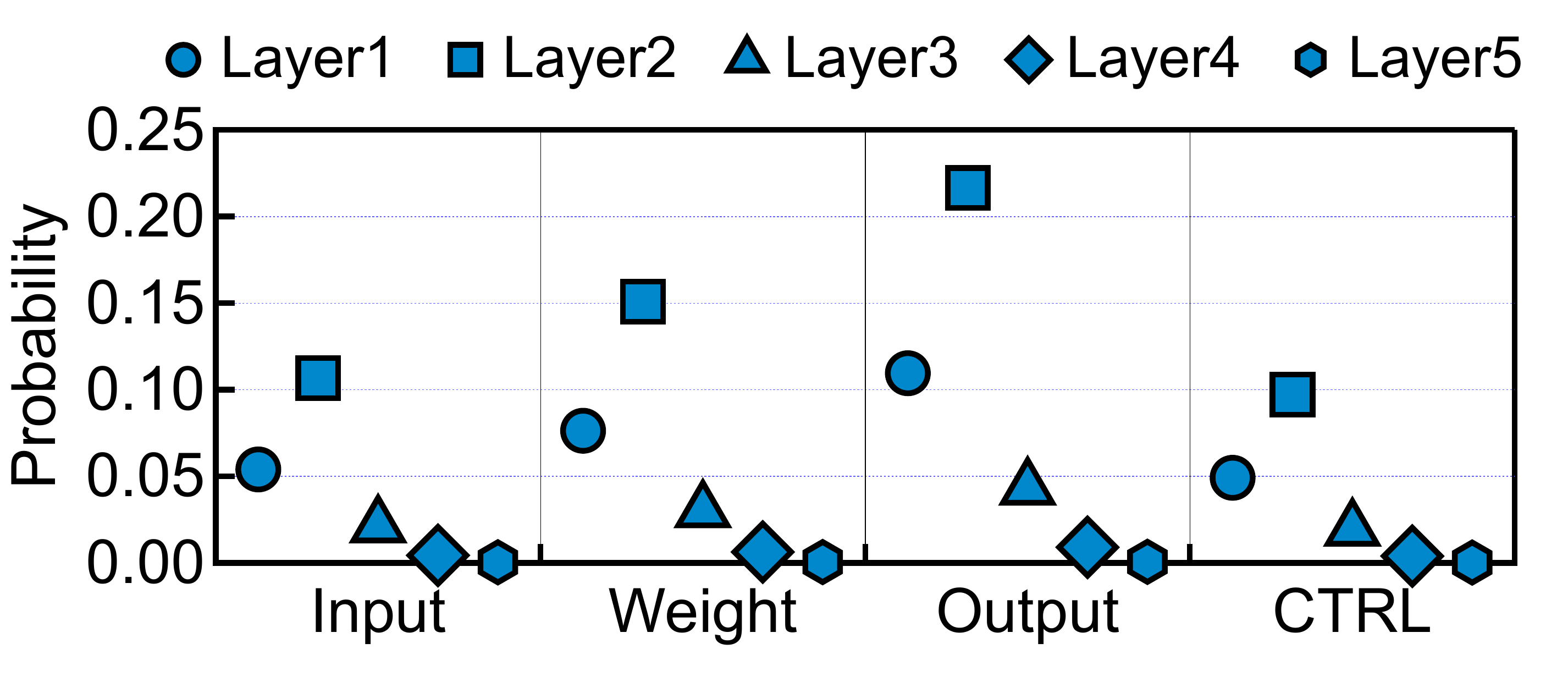}
  \label{fig:probra_lenet}
}
\hspace{2pt}
\subfloat[\small{MNIST-Hogwild.}]
{
  \includegraphics[trim=0 0 0 0, clip, width=0.9\columnwidth]{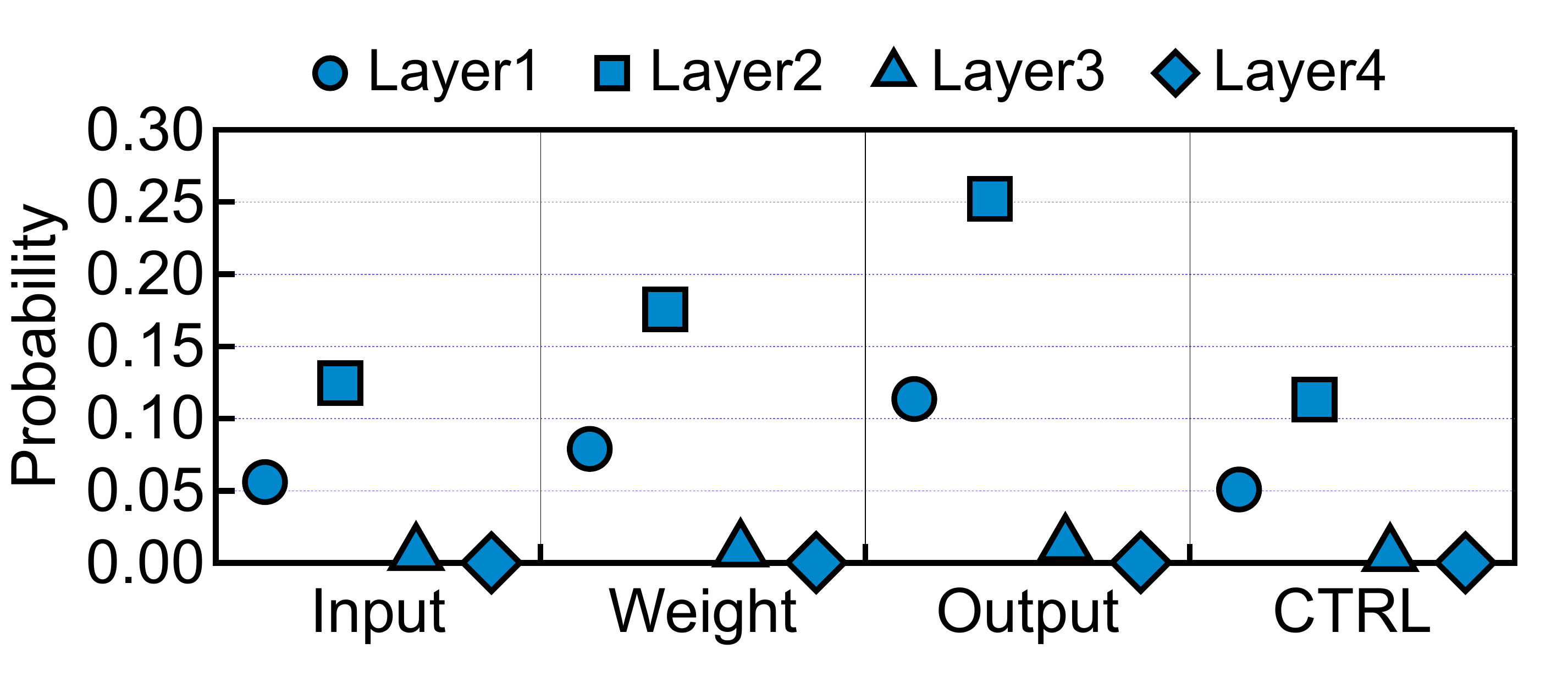}
  \label{fig:probra_simplenet}
}
\caption{The probability a transient fault is received by an input activation, a weight, an output activation, or a control variable across layers in LeNet-5 and MNIST-Hogwild. Software variables do not have uniform faulty probabilities. Control variables do not correspond to variables in a DNN model --- they are proxies for modeling control FFs; see \Sect{sec:robust:def} for details.}
\label{fig:probra_moti}
\end{figure*}

\subsection{Existing RA Formulation}
\label{sec:motivation:sw}

In theory, the RA is defined at the hardware level:
\begin{align}
    RA = \frac{1}{N}\sum_{i=1}^{N}A(i)
    \label{eq:RA}
\end{align}
\noindent where $N$ is the total number of FFs in the hardware, and $A(i)$ is the resulting inference accuracy under a fault at the hardware fault site $i$. This equation is infeasible to use in practice due to the need to perform time-consuming RTL fault injections and simulations, which are four to five orders of magnitude slower than an inference in software.
Nevertheless, \Equ{eq:RA} expresses the ground truth RA of a DNN on a given hardware and serves as a reference that other more lightweight methods should strive to match.

Due to the infeasibility of performing RTL simulations, a common formulation of RA as used in today's DNN resiliency analysis tools such as PyTorchFI~\cite{mahmoud2020pytorchfi} and TensorFI~\cite{chen2020tensorfi} is to model soft errors directly in a DNN model.
In this method, a fault site is a bit position (BP) of a weight or an activation. Thus, the total number of fault sites is:
$$ \widehat{N_{fs}} = (\# of~weights + \# of~activations) \times \# of~BPs$$


Given the software faults sites, RA is estimated by 1) first uniformly-at-random sampling of the software fault sites and, for each sampled fault site, perform an fault injection to obtain the corresponding inference accuracy under that fault injection, and then 2) averaging the resulting inference accuracies across fault injections\footnote{Section IV-A in PyTorchFI~\cite{mahmoud2020pytorchfi}: ``\textit{In each inference run, we inject a single-bit flip in a randomly selected neuron in the DNN to emulate a computational hardware error that may occur during inference.}''. Section IV-A in TensorFI~\cite{chen2020tensorfi}: ``\textit{we perform 1000 random FI experiments per fault configuration and input.}''}. This is equivalent to calculating RA by:
\begin{align}
    RA_{SW} = \frac{1}{\widehat{N_{fs}}}\sum_{j=1}^{\widehat{N_{fs}}}A(j)
    \label{eq:rasw}
\end{align}

\noindent where $A(j)$ is the inference accuracy under a fault at a software fault site $j$. $A(j)$ is estimated through software fault injection: flipping the bit corresponding to the software fault site $j$ and performing an inference.

\subsection{Error Sources in Existing RA Formulation}
\label{sec:motivation:errors}



Existing RA formulation is flawed, because of incorrect assumptions made in both terms in \Equ{eq:rasw}.
The issue with the existing RA formulation is a DNN-specific instance of the general AVF-vs-SVF mismatch as analyzed in detail by Papadimitriou and Gizopoulos~\mbox{\cite{papadimitriou2021demystifying}}. Let us elaborate this mismatch in the DNN context.

\paragraph{Inaccurate Inference Accuracy.} 
Estimating $A(j)$ using software fault injection does not reflect accuracy impact of actual soft errors in hardware --- for two reasons.

First, software fault injection assumes that, based on the data dependencies, one bit flip in a kernel weight corrupts \textit{all} the output activations.
However, the actual number of corrupted activations depends on the reuse factor of the FF that holds the weight: if the weight FF is reused, say, 4 times before a new weight is written to the FF (e.g., due to the scheduling algorithm), the original weight affects only 4 output activations.
The same argument applies to activations.

Second, software faults considered in existing tools do not reflect hardware fault sites.
For instance, these tools make no distinction between an output activation and its corresponding input activation in the next layer (because numerically they are the same value), but input activations and output activations are stored in two different sets of hardware FFs that might have different faulty probabilities.
Software fault injection also does not consider the effects of global \textit{control FFs}, whose bits when corrupted drastically reduce accuracy. When transient faults take place in control FFs, the accelerator usually crashes, severely degrading the inference accuracy~\cite{he2020fidelity}.

FIdelity~\cite{he2020fidelity} establishes a RTL-validated method that addresses the two issues above and, thus, correctly obtains the accurate $A(j)$ from software fault injection. Our work builds on FIdelity to obtain an accurate $A(j)$ estimation when needed, and does \textit{not} claim it as our contribution.

\paragraph{Inaccurate Fault Site Probability.} By simply averaging the inference accuracy across all the software fault injections, the RA formulation in \Equ{eq:rasw} implicitly assumes that all software fault sites are equally probable to transient errors. This assumption, however, is incorrect, because of how software variables are stored and reused in hardware FFs.


To demonstrate this
\Fig{fig:probra_moti} shows the faulty probability of software fault sites in two DNNs, LeNet-5~\cite{deng2012mnist} and MNIST-Hogwild~\cite{NEURIPS2019_9015}, running on NVDLA.
Effectively, we uniformly inject transient faults to hardware and analyze which weights/activations are effected. In the end, we aggregate statistics of the faulty probability of software fault sites.
We assume a raw FIT rate of 600/MB following the study from Jagannathan et al.~\cite{jagannathan2012frequency} and consistent with FIdelity~\cite{he2020fidelity}.

The probabilities of software fault sites vary significantly, both across variable type (e.g., input activations vs. weights) and across layers.
For instance, the faulty probabilities of variables in the last layer are two orders of magnitude lower than those in earlier layers.
This is because weights/activations in layers that take longer time to execute stay in the hardware longer and, thus, are more vulnerable to transient errors.

In addition, \Fig{fig:probra_moti} also shows the faulty probability of control variables, which do not correspond to any real variables in a DNN model; rather, they are proxies used to capture control FFs in hardware (see \Sect{sec:robust:def} for details). Those control variables generally have lower faulty probabilities than weights/activations, because there are fewer control FFs than FFs for weights/activations. As a result, they are less likely to receive transient faults than FFs of other categories.



%% file: robust.tex
\section{Defining and Modeling RA}
\label{sec:ptf}

We first introduce our RA formulation (\Sect{sec:robust:def}), followed by an algorithm that calculates the faulty probability of any software fault site (\Sect{sec:robust:algo})

\subsection{Defining RA as an Expected Value}
\label{sec:robust:def} 

Recall the intuition behind RA: it should capture the network inference accuracy \textit{given} that a fault has occurred.
Borrowing ideas from probability theory, we define RA as the \textit{expected value}~\cite{ev} of a network's inference accuracy under faults. Following the definition of the expected value, the equation to calculate RA takes the average of the network inference accuracy under each software fault \textit{weighted} by the faulty probability of each software fault site:
\begin{align}
    RA = \sum_{j=1}^{N_{fs}}p(j)A(j)
    \label{eq:rahw}
\end{align}

\noindent where $N_{fs}$ denotes the total number of software fault sites, $p(j)$ denotes the probability that a software fault site $j$ experiences a transient error, and $A(j)$ denotes the inference accuracy given the transient error at fault site $j$ and can be calculated using the FIdelity framework~\cite{he2020fidelity}.

The intuition behind our formulation is that a network's RA is a random variable, whose outcome depends on a large number of independent events, i.e., transient faults. Each transient fault affects one software fault site $j$ (e.g., one bit flip in a weight FF affects a weight in the model) with a different probability $p(j)$ and results in an inference accuracy $A(j)$. 
Note that the expected value equation inherently iterates over all software fault sites, but this is different from saying that each inference will actually have a transient fault in reality.

Our RA metric is nothing more than the \mbox{\textit{DNN-specific AVF}}.
For CPUs and GPUs, AVF is defined as the probability (0 or 1) that a transient fault in the hardware leads to an error in the application; the so-derived AVFs of all bits and cycles are averaged to get the overal AVF.
The difference in DNNs is that the AVF of a single bit in a single cycle is the DNN inference accuracy under that fault and, thus, can be any value between 0 and 1, which is captured by our $A(j)$. In addition, since we aim to estimate DNN RA/AVF in software, the fault sites are defined at the software level, which we elaborate next.

\no{While the formulation in \Equ{eq:rahw} appears very similar to the existing RA formulation in \Equ{eq:rasw}, there are two differences. First, we do not assume that software fault sites have equal faulty probability. In fact, each software fault site $j$ has a faulty probability $p(j)$ that we will analytical derive later. Second, the software fault sites we consider match closely with how faults are received in hardware. We elaborate on this below.}


\paragraph{Software Fault Sites.} A software fault site is a 2-tuple $<\mathbf{Var, BP}>$, where $Var$ is a software variable that can potentially become faulty due to a transient error and $BP$ is a bit position in a $Var$.
In particular, there are four types of software variables we model: 1) input activations, 2) output activations, 3) weights, and 4) control variables. Note that while a layer's output activations are the next layer's input activations, they must be separately modeled because they are held in different FFs in common DNN accelerators and, thus, are independently vulnerable to transient errors. This is one key difference compared to software fault sites considered in the existing RA formulation in \Equ{eq:rasw}.

Another key difference is control variables, which do not correspond to any real variable in a DNN model; rather, they are a proxy to capture FFs that store neither weights nor activations. Control FFs fall into two categories: \textit{global} control FFs that are used for data sequencing and address generation and \textit{local} control FFs that are tightly coupled with the datapath (e.g., valid bit, MUX select bits).
Faults in global control FFs generally crash the accelerator, providing no inference result for a particular input. Local control flops affect neurons depending on where they are located. In our model, we assign a software variable to each control FF in hardware.


For simplicity purposes, we omit $BP$ in later discussions and equate a software fault site with a software variable. The faulty probabilities so derived must be multiplied by $\frac{1}{BP}$ to obtained the actual probabilities of software fault sites, given the assumption that the probability of the bit flip occurring at a specific bit is the same across all bit positions in a FF.


\subsection{Calculating RA Through Probability Transfer}
\label{sec:robust:algo}

We present an algorithm to estimate the transient error probability of software fault sites; it works by transferring the faulty probability from the hardware to the software.

\begin{figure}[t]
  \centering
  \includegraphics[width=\columnwidth]{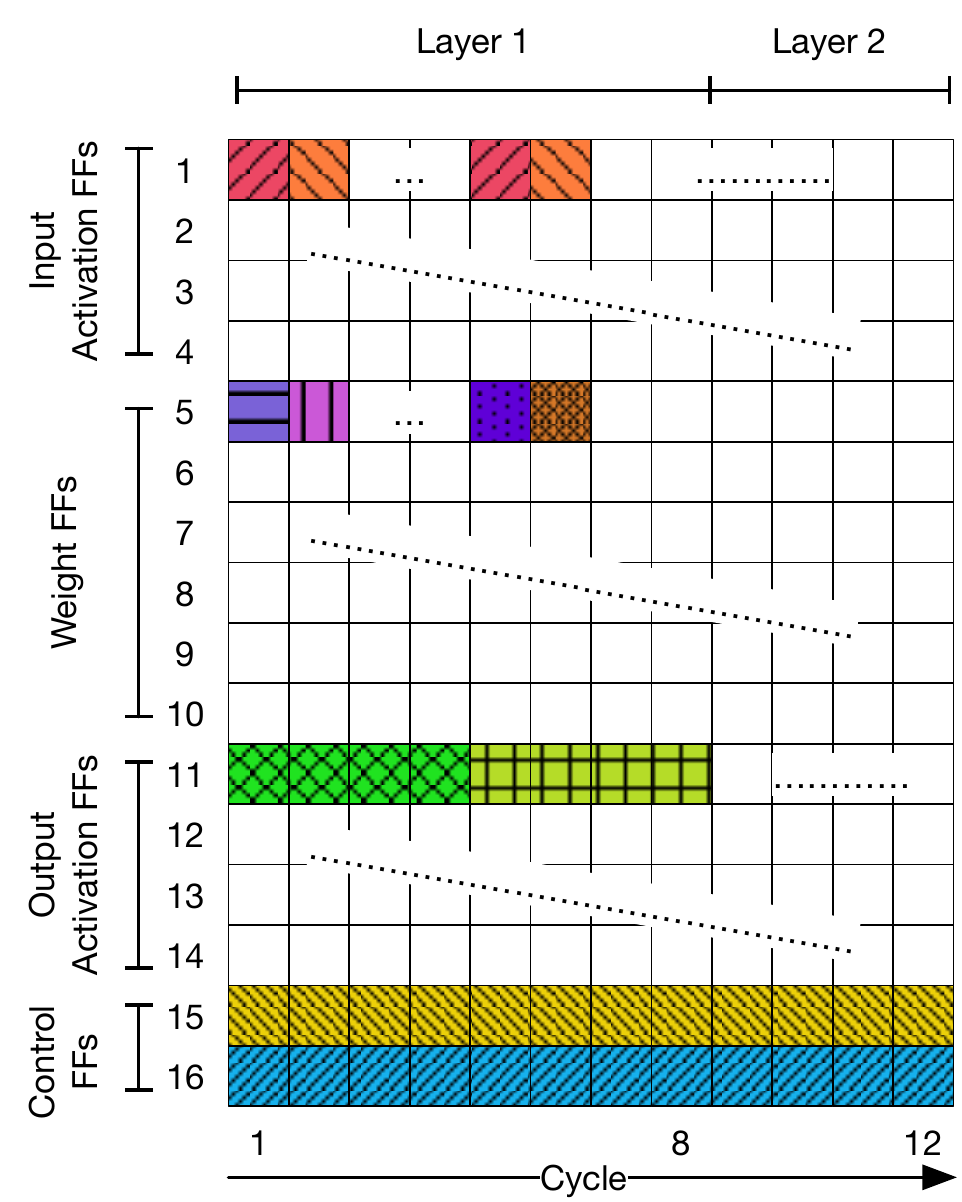}
  \caption{An (not-to-scale) illustration of the idea of calculating the faulty probability of a software fault site. Each cell represents a hardware fault site (i.e., a particular FF at a particular cycle). Each stripe pattern (color) represents a software variable, which can occupy multiple cells because of data reuse. The goal is to calculate the number of cells each software variable is mapped to.}
  \label{fig:fault}
\end{figure}

\paragraph{Intuition.} Our idea is illustrated in \Fig{fig:fault}, where each row corresponds to a hardware FF and each column, from left to right, represents a cycle over time. Thus, each cell represents a unique hardware fault site (i.e., a particular FF at a particular cycle).
A transient error can be thought of as dropping a pin on the 2D grid. The probability of a pin landing at a cell is dictated by the raw FIT rate of the corresponding FF. For instance, if the raw FIT rates for hardware FFs are uniform, the pin has an equal probability of landing at any cell.

The central question is, what are the probabilities of a randomly dropped pin landing on \textit{each of the software fault sites}?
The crux is to model \textit{how software fault sites are mapped to the cells}.
A useful thought experiment is that if a software fault site $j$ is mapped to $T(j)$ cells out of a total amount of $M$ cells, and each cell has the same probability of receiving the pin (e.g., if hardware FFs have the same raw FIT rate), the probability that $j$ receives a transient error is $\frac{T(j)}{M}$. This can be easily generalized to cases where hardware FFs do not have the same raw FIT rate as we will show later.

Note that each software variable occupies multiple cells because of data reuse. For instance, in a weight stationary accelerator a weight will stay in a FF for $C$ cycles and reused by $C$ input activations (the exact number of $C$ depends on the scheduling/tiling policy). As a result, that weight will occupy $C$ consecutive cells belonging to the holding FF. The same argument applies to activations, too. A control variable stays in a control FF throughout the execution, because each control variable is uniquely associated with a control FF.

\paragraph{Assumptions.} To model the mapping, we make the following assumptions about DNN execution generally true to common DNN accelerators:

\vspace{2pt}
\noindent\mode{Assumption 1} A DNN is executed layer by layer.

\vspace{2pt}
\noindent\mode{Assumption 2} Each software variable resides in only one type of FFs. For instance, a weight is always found in a weight FF (although it could be found in different weight FFs at different cycles) and is not in an activation FF. In other words, a FF will not hold software variables of different types.
While perhaps obvious, we perform a large-scale RTL fault injection (over 2.6 billion in total) on a systolic array accelerator, and inspect the resulting Fast Signal DataBase (FSDB) waveforms, to confirm this. See \Sect{sec:val} for the setup.

The implication is that, \textit{given} that a transient fault occurs, the probability that the fault occurs to a particular FF type is equivalent to the percentage of the FFs of that type.

\vspace{2pt}
\noindent\mode{Assumption 3} Different software variables of the same type spend the same amount of time in a FF. For instance, a weight FF throughout the execution will hold different weights, which spend the same number of cycles in that FF. This is fundamentally true since variables of the same type have the same number of reuses in a DNN.   
Again, from the over 2.6 billion RTL fault injections and simulations above we inspect the FSDB waveforms to confirm this. Across 200,736 weights of a layer, the standard deviation of the number of cycles a weight spent on a weight FF is 0.025 (not exactly 0 due to the ramp-down in the end).
    
The implication is that \textit{given} that a transient fault occurs in a weight (activation) FF, all the weights (activations) in the same layer are equally probable of receiving that fault.

\vspace{2pt}
\noindent\mode{Assumption 4} The accelerator utilization is near 100\%. That is, every cell is associated with a software fault site throughout the execution (no ``empty cell''). This assumption does not hold in extreme cases where the hardware resources are abundant but the network is extremely small (e.g., a 1K $\times$ 1K PE array executing a layer where the feature map dimension is 4 $\times$ 4).
At the end of this section, we will relax this constraint and show how RA is estimated with an actual utilization factor.
    

\paragraph{Derivation.} Given the observations above, the probability of a software fault site $j$ can be expressed as:
\begin{align}
    p(j) &= T(j) \times p_{T_j}
    \label{eq:pequ}
\end{align}

\noindent where $T(j)$ denotes the number of cells that a software fault site $j$ (of type $T_j$ in layer $L_j$) occupies, and $p_{T_j}$ denotes the probability that a cell of type $T_j$ receives a transient fault.

$T(j)$ is further modeled as:
\begin{align}
    T(j) = M \times LP(L_j) \times TP(T_j) \times VP(j)
    \label{eq:raequ}
\end{align}

\noindent where:

\begin{itemize}
    \item $M$ is the total number of cells, which is 192 in \Fig{fig:fault}.
    \item $LP(L_j)$ is the percentage of the cells used by layer $L_j$. In the example of \Fig{fig:fault}, if $L_j$ is the first layer, $LP(L_j)$ would be $\frac{2}{3}$. Thus, $M \times LP(L_j)$ is the total number of cells belonging to layer $L_j$.
    \item $TP(T_j)$ denotes the percentage of the FFs of type $T_j$. It is equivalent to the percentage of the cells that belong to type $T_j$. In the example of \Fig{fig:fault}, if $T_j$ is the input activation type, $TP(T_j)$ would be $\frac{1}{4}$. Thus, $M \times LP(L_j) \times TP(T_j)$ is the total number of cells belonging to FF type $T_j$ in layer $L_j$.
    \item $VP(j)$ is the probability that a transient fault occurs to a variable $j$ given that a fault has occurred to some variable of type $T_j$. Thus, $M \times LP(L_j) \times TP(T_j) \times VP(j)$ is the total number of cells belonging to the variable $j$.
\end{itemize}

Taking another perspective, \Equ{eq:raequ} essentially computes the joint probability of three independent events: 1) a fault occurs while layer $L_j$ is executing; the probability of this event is $LP(L_j)$, 2) a fault occurs in an FF that is of type $T_j$; the probability of this event is $TP(T_j)$, and 3) a fault occurs to a particular variable $j$ among all the variables of the type $T(j)$; the probability of this event is $VP(j)$.
Let us now analytically derive each of these terms in \Equ{eq:raequ}.


$LP(L_j)$, given \mode{Assumption 1}, is equal to the percentage of the execution time layer $L_j$ takes, which, given \mode{Assumption 4} (the hardware utilization is nearly full), can be modeled as the ratio between the Multiply and Accumulate (MAC) operation count in that layer over the total number of MAC operations in the network:
\begin{align}
	LP(L_j) = \frac{MAC(L_j)}{\sum_{i=1}^{UF(j) + p(j)SA(1-UF(j))L}MAC(L_i)}
	\label{eq:LayerProb}
\end{align}

\noindent where $MAC(L_j)$ denotes the number of MAC operations performed in layer $L_j$, and $L$ is the total number of layers.

$TP(T_j)$ denotes the percentage of the FFs of type $T_j$. Given \mode{Assumption 2}, $TP(T_j)$ is given as:
\begin{align}
	TP(T_j) = \frac{FF(T_j)}{\sum_{i=1}^{T}FF(T_i)}
	\label{eq:TypeProb}
\end{align}

\noindent where $FF(T_j)$ denotes the number of FFs of type $T_j$, and $T$ is the total number of FF types.

Based on \mode{Assumption 3} that all variables of the same type are equally probable of receiving a fault, $VP(j)$ is equal to the probability that a variable $j$ is selected out of all the variables of the same type $T_j$, and can be expressed as:
\begin{align}
    VP(j) = \frac{1}{V(T_j)}
\end{align}

\noindent where $V(T_j)$ is the total number of variables of type $T_j$. For control variables, $V(T_j)$ is always 1 as each control FF is mapped to one control variable.

$p_{T_j}$, the probability of a cell of type $T_j$ receiving a transient fault, is given by:

\begin{align}
    p_{T_j} = \frac{rawFIT(T_j)}{M \times \sum_{i=1}^{T}(TP(T_i)\times rawFIT(T_i))}
    \label{eq:pcell}
\end{align}

\noindent where $rawFIT(T_j)$ denotes the raw FIT rate of FFs of type $T_j$. One way to interpret \Equ{eq:pcell} is that if all the FFs have the same raw FIT rate, $p_{T_j}$ is reduced to $\frac{1}{M}$, matching our intuition that when all FFs are equally faulty each cell has the same chance of receiving the fault.

Technically, the traditional AVF metric is not concerned with the raw FIT rate, which, among other factors, could vary between different FF types (e.g., different sizes for different FF types). We include the raw FIT rate in the $p_{T_j}$ formulation just to show how to incorporate the impact of different raw FIT rates, as we will later show in \mbox{\Fig{fig:fit}}. To calculate the actual AVF, $p_{T_j}$ should simply be $\frac{1}{M}$.

Putting it together, \Equ{eq:pequ} requires only basic statistics of the hardware accelerator and the DNN, and is extremely lightweight to calculate. Thus, one could calculate the fault probability for all the software fault sites, even if there are billions of them. Our approach can be seen as transferring the hardware fault probability (raw FIT rate) to software fault probability, hence the name ``probability transfer.''


\paragraph{Using Actual Utilization Factor.} So far we have assumed that the \textit{Utilization Factor} $UF(j)$ of all the cells in \Fig{fig:fault} is 100\%, which is obviously not always true. When a cell is not utilized, i.e., a FF does not hold an actual value pertaining to inference, a soft error occurring to that FF has no impact on the inference accuracy.
While UF in theory is a hardware concept, statistically speaking the UF of a hardware FF type is the same as the UF of the corresponding software variables. This is because each software variables of the same type is assigned the same amount of cells/FFs (\mode{Assumption 3}).
Therefore, RA can be formulated as the average of the network inference accuracy under each software fault weighted by both the faulty probability of each software fault site and the \textit{Utilization Factor} $UF(j)$ of a software fault site $j$:

\begin{align}
    RA = \sum_{j=1}^{N_{fs}}(p(j)UF(j)A(j) + p(j)(1-UF(j))SA)
    \label{eq:rahw_uf}
\end{align}

\noindent where ($SA$) is the fault-free, standard inference accuracy of the network.
Intuitively, \Equ{eq:rahw_uf} says a software fault site $j$ can affect the inference accuracy for just $UF(j)$ portion of the time; for the remaining $(1-UF(j))$ portion, it will have no impact on the overall accuracy of the network, resulting in an $A(j)$ equal to the SA of the network. We will show in \Sect{sec:val:prob} that using an actual UF obtained from RTL simulation improves the RA estimation accuracy.

%% file: sample.tex
\section{Estimating RA Using Importance Sampling}
\label{sec:sampling}

To accelerate RA convergence, we propose to formulate RA estimation as a Monte Carlo integration problem and solve it using importance sampling (\Sect{sec:sampling:mc}). We show that perfect importance sampling is fundamentally impossible, and propose two heuristics that leverage DNN-specific characteristics to approximate importance sampling while retaining the main benefits (\Sect{sec:sampling:is}).

\subsection{Compute RA via Monte Carlo Integration}
\label{sec:sampling:mc}

\Equ{eq:rahw} presents an analytical form of accurately calculating the RA of a DNN. However, that equation is impractical to calculate because of the large $N$ number, i.e., the number of software fault sites. While $p(j)$ for each fault site $j$ can be trivially calculated using \Equ{eq:pequ}, obtaining $A(j)$ requires inferencing over an entire test set. Take MobileNet as an example: it has about 3.5 billion software fault sites; assuming obtaining $A(j)$ for each fault site requires 1 second, it would take 111 years to precisely calculate the RA for MobileNet using the exact formulation in \Equ{eq:rahw}.

Our observation is that calculating RA can be seen as integrating a discrete function over a finite domain:
\begin{align}
    RA = \sum_{j=1}^{N}f(j)\\
    f(j) = p(j)A(j),~~~j \in \mathbb{Z} : j \in [1, N]
\end{align}

The integrand $f(\cdot)$ does not have an analytical form that can be calculated in practice. We propose to solve the integration numerically using Monte Carlo integration~\cite{press2007numerical}.

Formally, RA can be estimated by drawing K independent samples $X_1, ..., X_K$ using a probability density function (PDF) and calculate:
\begin{align}
    \widehat{RA} =  \frac{1}{K}\sum_{j=1}^{K}\frac{f(X_j)}{PDF(X_j)},~~~\sum_{j=1}^{N}PDF(X_j) = 1
\end{align}

\begin{figure}[t]
\centering
\subfloat[Uniform sampling (constant PDF).]{
    \label{fig:unisamp}
    \includegraphics[width=0.465\columnwidth]{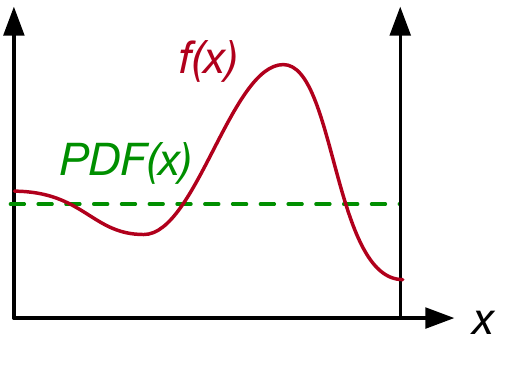}
}
\hfill
\subfloat[Importance sampling, where the PDF is proportional to integrand $f(\cdot)$.]{
    \label{fig:impsamp}
    \includegraphics[width=0.465\columnwidth]{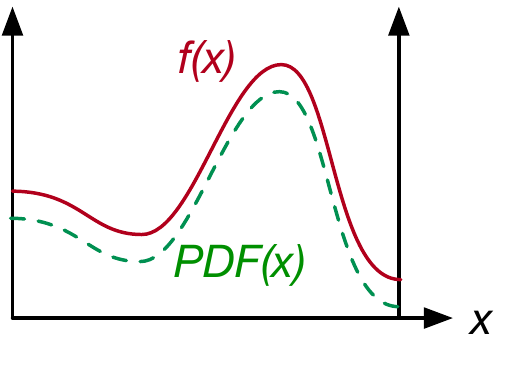}
} 
\caption{Comparison of two PDFs used in sampling. Note that the PDF must integrate to 1 by definition.}
\label{fig:samp}
\end{figure}

$\widehat{RA}$ is called the Monte Carlo estimator of RA. It can be shown that when K approaches infinity $\widehat{RA}$ converges to $RA$ (or alternatively, the expected value of $\widehat{RA}$ is $RA$).
However, different sampling methods have significantly different convergence rate.
A naive method is to sample the domain uniformly as shown in \Fig{fig:unisamp}, essentially using a PDF of $\frac{1}{N}$.
While a common choice, uniform sampling takes a lot of samples to converge as we will show later\footnote{More formally, the standard deviation of uniform sampling is proportional to $\frac{1}{\sqrt{N}}$, indicating that we must quadruple the number of samples to reduce the error by half.}.

\paragraph{Importance Sampling.} To improve the convergence rate, an observation is that we have the freedom to choose the PDF used to draw the samples. Intuitively, we want to place more samples where the contribution to the integration is numerically high (i.e., ``important''), hence capturing most of the integral contributions. \Fig{fig:impsamp} illustrates the idea.

In particular, it is established that if the $PDF(\cdot)$ is exactly proportional to the integrand $f(\cdot)$, that is:
\begin{align}
    PDF(j) = c f(j),
\end{align}

\noindent where $c$ is a normalization value that guarantees that $PDF(j)$ integrates to 1:
\begin{align}
    c = \frac{1}{\sum_{j=1}^{N}f(j)}
\end{align}

\noindent then the variance of the Monte Carlo estimator is zero, indicating that exactly \textit{one} sample is sufficient~\cite{kroese2013handbook}.

However, perfect important sampling is impossible, because determining $c$ requires us to calculate the integration that we set out to estimate in the first place.

\subsection{Heuristics for Importance Sampling}
\label{sec:sampling:is}

We do not know $f(\cdot)$ precisely and, thus, cannot construct a PDF exactly proportional to $f(\cdot)$. However, we observe that $f(\cdot)$ is a product of two terms: $p(\cdot)$ and $A(\cdot)$. Our idea is to use \textit{a priori} knowledge of $p(\cdot)$ and $A(\cdot)$ to construct a PDF that is \textit{approximately} proportional to $f(\cdot)$.

In particular, $p(\cdot)$ can be completely constructed in advance given a particular DNN and the underlying hardware using \Equ{eq:pequ}. In contrast, $A(\cdot)$ cannot be constructed offline in practical terms since it requires enumerating hundreds of billions of fault sites, each of which requires performing inference on an entire test set.

We propose two heuristics to empirically model $A(\cdot)$.
Each heuristic gives a unique approximation of $A(\cdot)$, which when multiplied with $p(\cdot)$ gives a $\hat{f}(\cdot)$, which is an approximation of $f(\cdot)$. We then construct the PDF proportional to $\hat{f}(j)$ for importance sampling.

\paragraph{Constancy Heuristic.} The simplest heuristic is to assume that $A(j)$ is constant. That is, the inference accuracy under any software fault site $j$ is the same. This heuristic in turn equates $f(\cdot)$ to $p(\cdot)$. Thus, the sampling PDF is constructed proportional to $p(\cdot)$.

\paragraph{BP Heuristic.} We can improve upon the uniform assumption by observing that transient errors in certain bit positions (BPs) have a higher impact on accuracy than other positions. Taking the floating point representation as an example, a bit flip in the sign and exponent fields changes the numerical values much more significantly than bit flips in the fraction field; as a result, prior work has shown that transient errors in the sign and exponent fields degrade the inference accuracy more than the fraction field~\cite{li2017understanding}.

Given this empirical observation, we can model $A(\cdot)$ with respect to the BP. The exact impact of each BP in a number representation can be profiled offline.

It is worth emphasizing that profiling does not have to (and will not) capture the exact impact of BPs on $A(\cdot)$. This is because $\hat{f}(\cdot)$, as an heuristic for importance sampling, does not have to be exactly the same as $f(\cdot)$. The approximation of $\hat{f}(\cdot)$ affects only the convergence rate, \textit{not} the accuracy of RA estimation.
We will show in \Sect{sec:ra} that even simple assumptions of the BP-wise impact are sufficient.


%% file: validation.tex
\section{Validation}
\label{sec:val}

\begin{figure*}[t]
\centering
\subfloat[\small{Input Activations.}]
{
  \includegraphics[trim=0 0 0 0, clip, height=0.40\columnwidth]{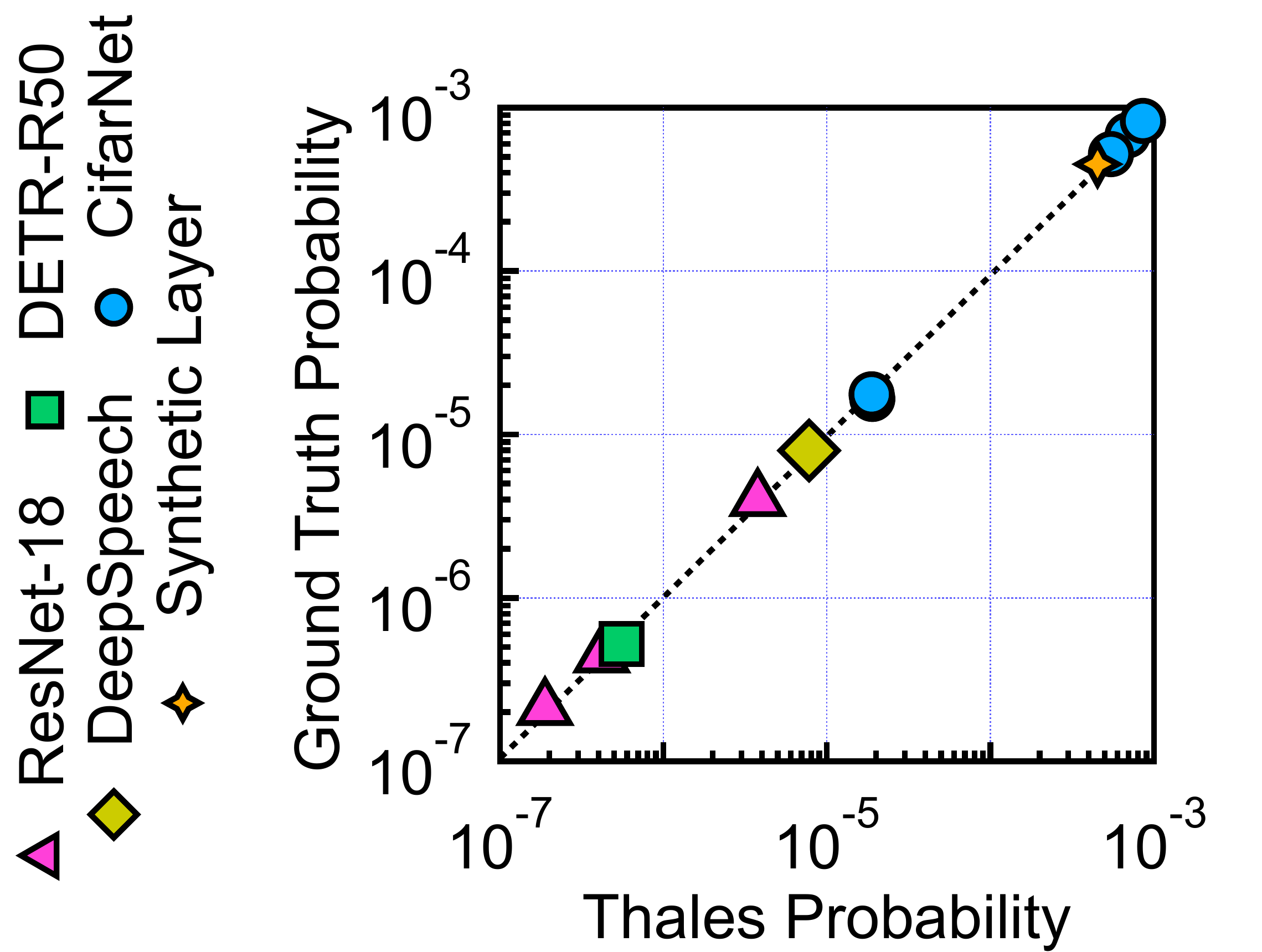}
  \label{fig:prob_IA}
}
\subfloat[\small{Weights.}]
{
  \includegraphics[trim=0 0 0 0, clip, height=0.4\columnwidth]{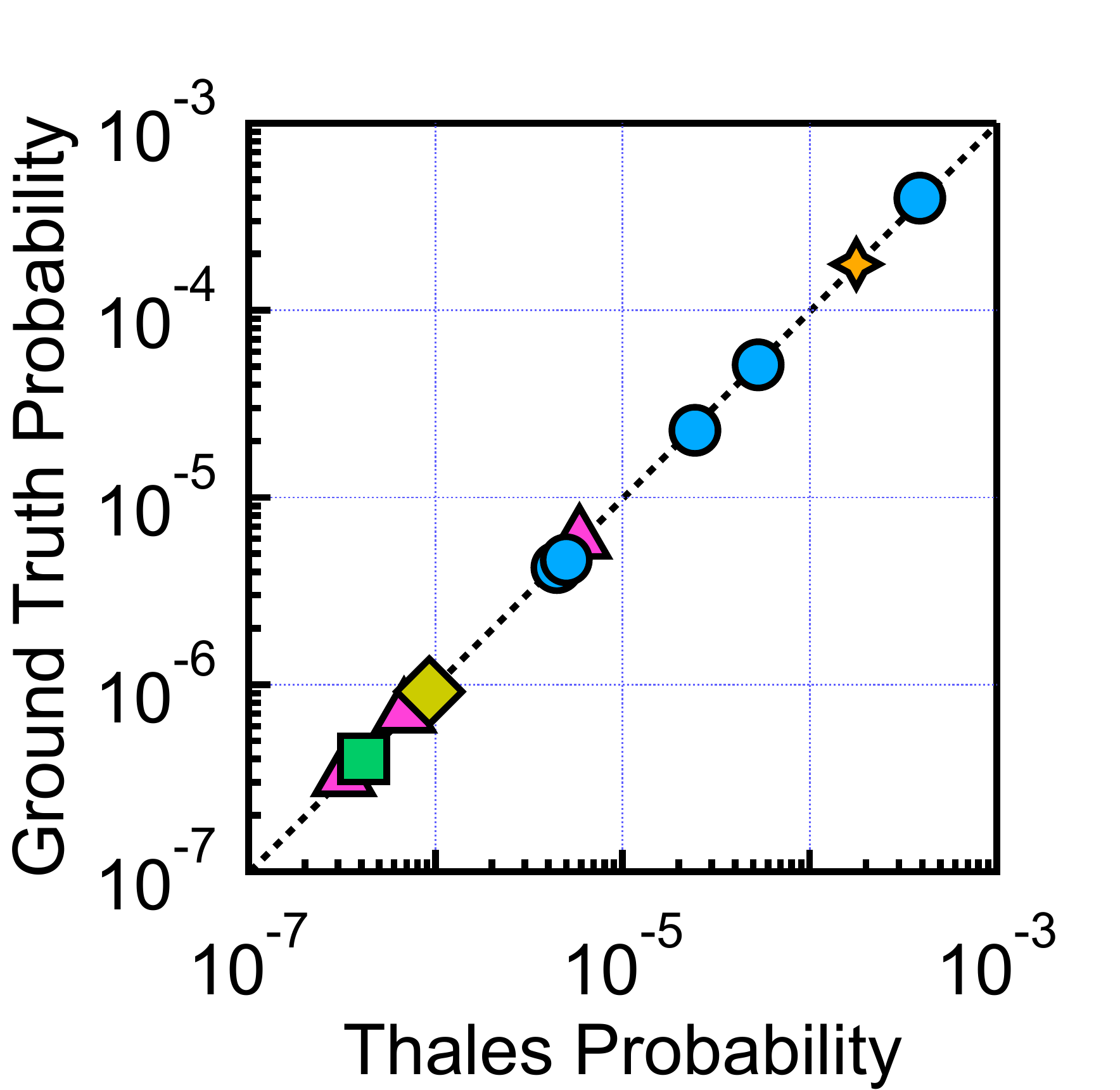}
  \label{fig:prob_W}
}
\subfloat[\small{Output Activations.}]
{
  \includegraphics[trim=0 0 0 0, clip, height=0.4\columnwidth]{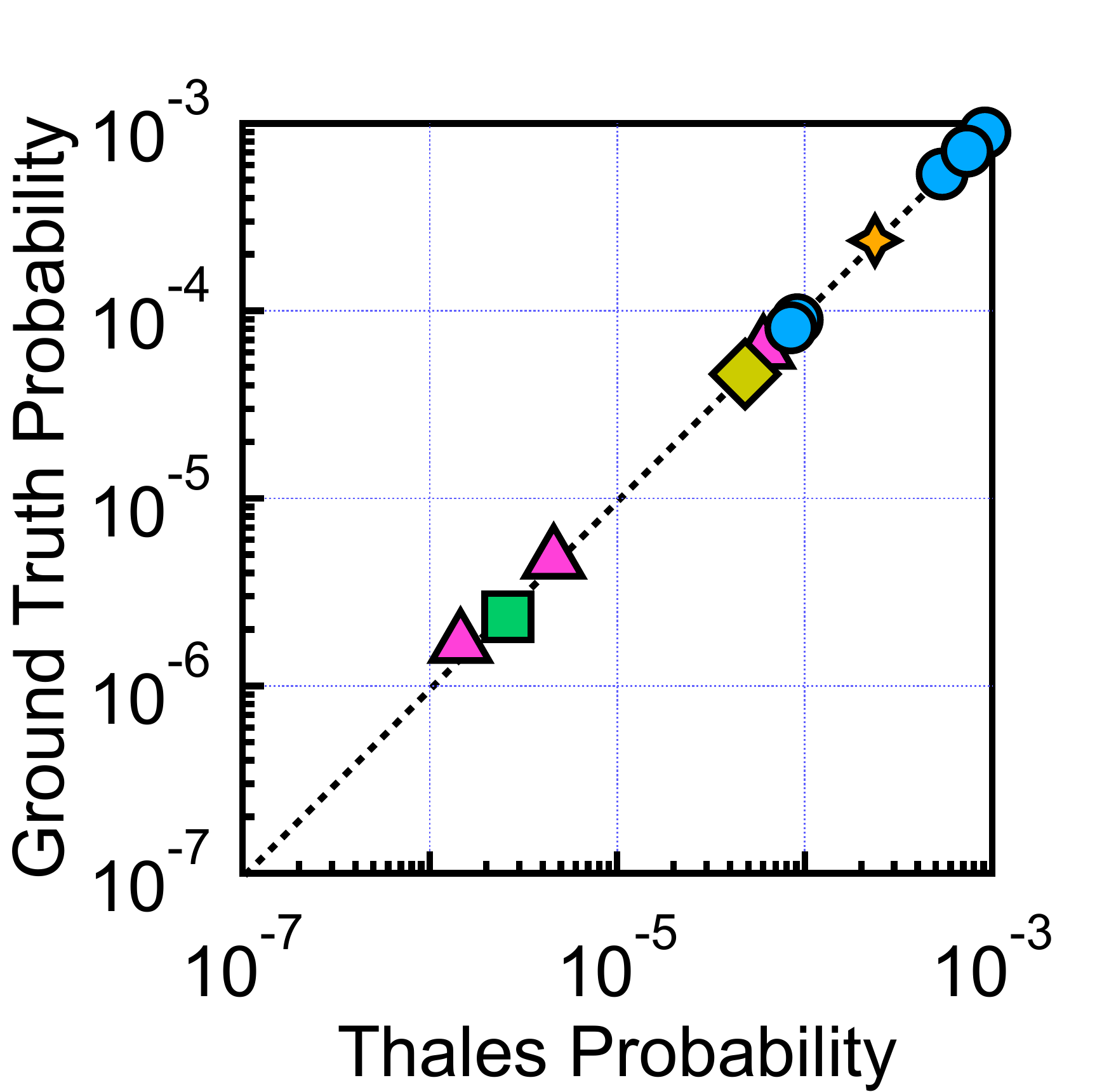}
  \label{fig:prob_OA}
}
\subfloat[\small{Control.}]
{
  \includegraphics[trim=0 0 0 0, clip, height=0.4\columnwidth]{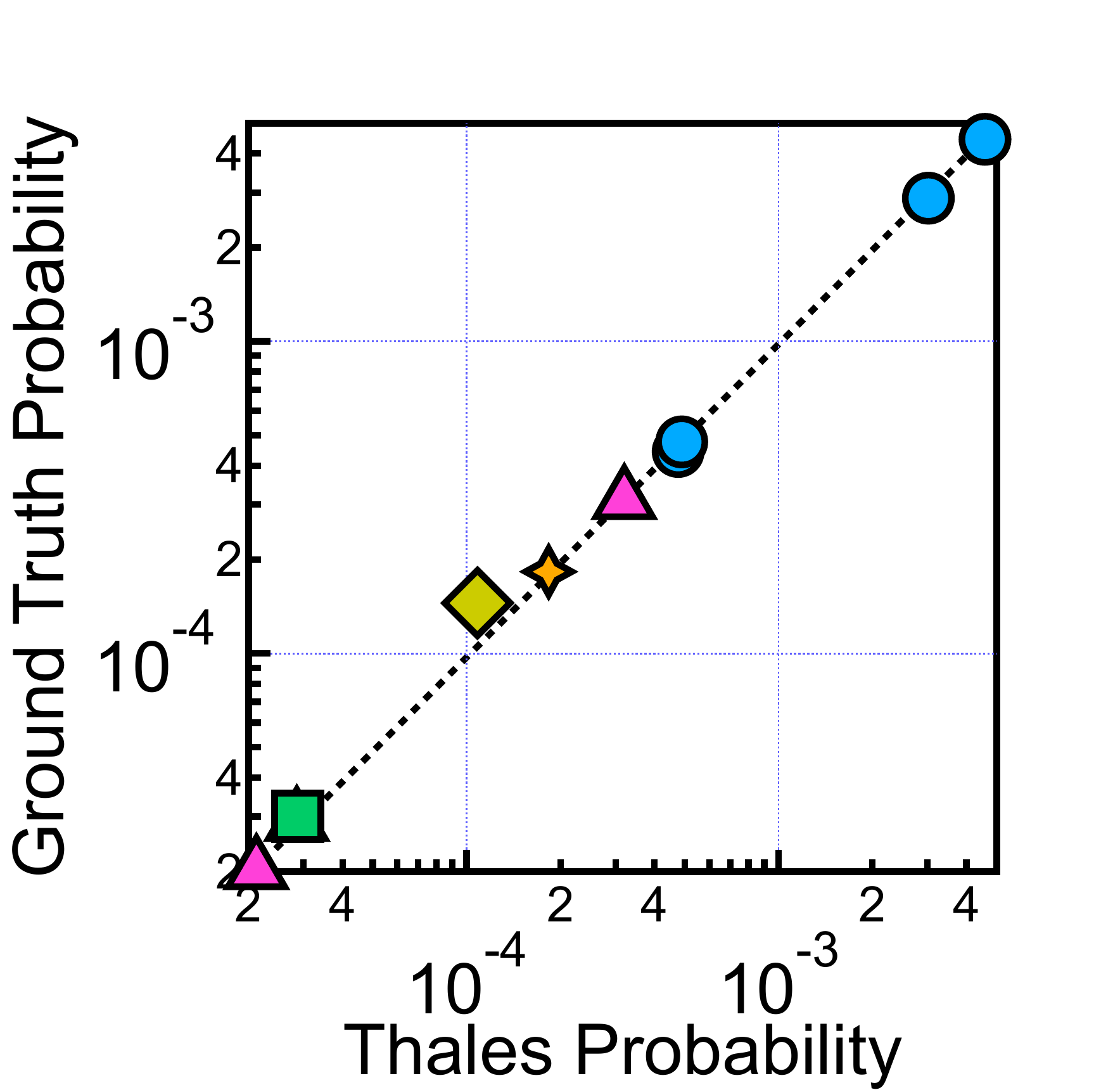}
  \label{fig:prob_cc}
}
\caption{Comparison of ground truth FF fault probabilities ($y$-axis) with analytically derived probabilities by \mbox{\proj} ($x$-axis).}
\vspace{-10pt}
\label{fig:prob_plot}
\end{figure*}

We first describe the RTL-based validation setup (\Sect{sec:val:setup}), and show that the software fault site probability and RA estimation analytically derived in \Sect{sec:robust:algo} match well with results obtained from RTL fault injection (\Sect{sec:val:prob}).

\subsection{Validation Setup}
\label{sec:val:setup}

We use a TPU-like systolic array with a 32 $\times$ 32 MAC array using the output stationary data-flow. After synthesis, the distribution across the input activation, weight, output activation, and control FFs is 30.56\%, 30.56\%, 21.87\%, 17.0\%.

\begin{table}[t]
\centering
\caption{Workloads for validation.}
\renewcommand*{\arraystretch}{1}
\renewcommand*{\tabcolsep}{4pt}
\resizebox{\columnwidth}{!}
{
\begin{tabular}{cccc}
\toprule[0.15em]
\textbf{Category} & \textbf{Network}  & \textbf{Layer(s)} & \textbf{Dataset}  \\
\midrule[0.05em]
- & Synthetic  & 7$\times$7 CONV & Random inputs\\
Classification       & CifarNet~\cite{hosang2015taking}  & All the layers  & CIFAR-10~\cite{krizhevsky2009learning}\\
Classification       & ResNet-18~\cite{he2016deep}  & 3x3 CONV (layer 2,4,6)  & ImageNet~\cite{deng2009imagenet}\\ 
Detection       & DETR-R50~\cite{carion2020end}  & 3x3 CONV (layer 7) & COCO~\cite{lin2014microsoft}\\ 
Speech-To-Text & DeepSpeech~\cite{hannun2014deep}  & FC (layer 10)  & LibriSpeech~\cite{panayotov2015librispeech}\\
\bottomrule[0.15em]
\end{tabular}
}
\vspace{-3mm}
\label{tbl:prob}
\end{table}

Similar to prior work on (DNN) resiliency~\mbox{\cite{he2020fidelity,lunardi2018efficacy,dos2021revealing}}, we use RTL fault injection to obtain ground truth data. We use Synopsys Z01X\mbox{\texttrademark~\cite{z01x}}, an industry-scale fault injection and simulation tool, to pick possible hardware fault sites written as $<\mathbf{Cycle, FF Type, BP}>$.
Each run, the tool injects a bit-flip injection at a fault site and performs the RTL simulation that takes the DNN inference to the end. 
Therefore, this methodology correctly captures the fact that one transient fault in a software variable (e.g., a weight) could result in errors in subsequent variables that depend on the variable receiving the initial transient fault (e.g., all the output activations).

\mbox{\Tbl{tbl:prob}} lists the workloads we use for validation. The first workload is a synthetic 7 $\times$ 7 CONV layer, for which we exhaust all the fault sites (i.e., every single bit at every single FF at every single cycle), resulting in a total of 2.629 Billion RTL-level fault injections and simulations (900 $\times$ 240 CPU hours in total).
The remaining workloads are individual layers from popular DNNs, for which we run fault injections and simulations on 332K randomly sampled fault sites, which are obtained using Synopsys Z01X to achieve a 99\% confidence level with a 1\% error margin.


\subsection{Validation Results}
\label{sec:val:prob}

\mbox{\paragraph{Fault Probability Validation.}} Our goal is to validate that the software fault site probability (i.e., $p(j)$ in the RA formulation in \mbox{\Equ{eq:rahw}}) is correctly calculated by our analytical model given in \mbox{\Equ{eq:pequ}}.
We take each sampled fault site and categorize them as either Input Activations, Weights, Output Activations or Control FFs. We then compare the faulty probability of each category that our model calculates with the data obtained from RTL fault injection and simulation.
We assume that the FFs have a raw FIT rate of 600/MB following the study from Jagannathan et al.\mbox{~\cite{jagannathan2012frequency}} and consistent with FIdelity\mbox{~\cite{he2020fidelity}}.

\mbox{\Fig{fig:prob_plot}} shows the validation results for the four FF categories.
Overall, the Pearson correlation coefficient is 0.9998 between results obtained by our analytical model and the ground truth, suggesting the accuracy of our approach. The main source of inaccuracy of is FC layers, where the analytically calculated probabilities are 6.4\% away from the ground truth. This is because FC layers do not fully utilize the hardware (74.0\% on average), while our model, absent detailed RTL-level data, assumes 100\% hardware utilization. When using the actual hardware utilization data, the estimation accuracy is brought to within 4.0\% of the ground truth.



\paragraph{RA Validation} We also validate that our RA estimation matches well with the ground truth RA ($RA_{True}$) obtained from RTL simulation.
To that end, we perform RTL fault injections and simulations on the entire CifarNet\mbox{~\cite{hosang2015taking}} and ResNet-18\mbox{~\cite{he2016deep}} performing inference on 20 randomly selected images from the CIFAR-10\mbox{~\cite{krizhevsky2009learning} dataset}. We obtain the Utilization Factor (UF) of both the networks from RTL simulations. We show two RA estimatios ($RA_{Est}$), first by assuming UF = 1 (i.e., using \Equ{eq:rahw}) and then by using the actual UF values (i.e., using \mbox{\Equ{eq:rahw_uf}}). The results are shown in \mbox{\Tbl{tbl:ravalidation}}.

We observe that the RA estimation is much closer to the ground truth RA when using the actual UF. Specifically, the RA when assuming UF = 1 is lower than that when using the actual UF. This is down to the fact that with higher utilization of the hardware, more FFs are vulnerable to soft errors. The percentage error in $RA_{Est}$ for CifarNet\mbox{~\cite{hosang2015taking}} and ResNet-18\mbox{~\cite{he2016deep}} reduces from 4.34\% and 6.35\% to 0.7\% and 1.58\% respectively when considering the actual UF.


\begin{table}[t]
\centering
\caption{RA estimation validation. $RA_{True}$ is the ground truth RA, $RA_{Est} $ is the RA estimated with Thales, and $UF$ is the utilization factor of the hardware obtained from RTL simulations. The error percentage is for $RA_{Est}$ estimated using the actual UF.}
\renewcommand*{\arraystretch}{1}
\renewcommand*{\tabcolsep}{4pt}
\resizebox{\columnwidth}{!}
{
\begin{tabular}{cccccc}
\toprule[0.15em]
\multirow{2}{*}{\textbf{Network}} & \textbf{UF} & $\mathbf{RA_{True}}$ & \multicolumn{2}{c}{$\mathbf{RA_{Est}}$}  & \textbf{\% Error}  \\
& &  & Actual UF & UF = 1 \\
\midrule[0.05em]
CifarNet~\cite{hosang2015taking} & 0.2632 & 0.930517  & 0.923981 & 0.890132 & 0.70240\\
ResNet-18~\cite{he2016deep}     & 0.36733 & 0.88445 & 0.870438 & 0.828221 & 1.5842\\
\bottomrule[0.15em]
\end{tabular}
}
\vspace{-3mm}
\label{tbl:ravalidation}
\end{table}


%% file: eval.tex
\section{Evaluating Sampling Strategies}
\label{sec:ra}

\begin{figure*}[t]
\centering
\subfloat[\small{LeNet.}]
{
  \includegraphics[trim=0 0 0 0, clip, width=0.40\columnwidth]{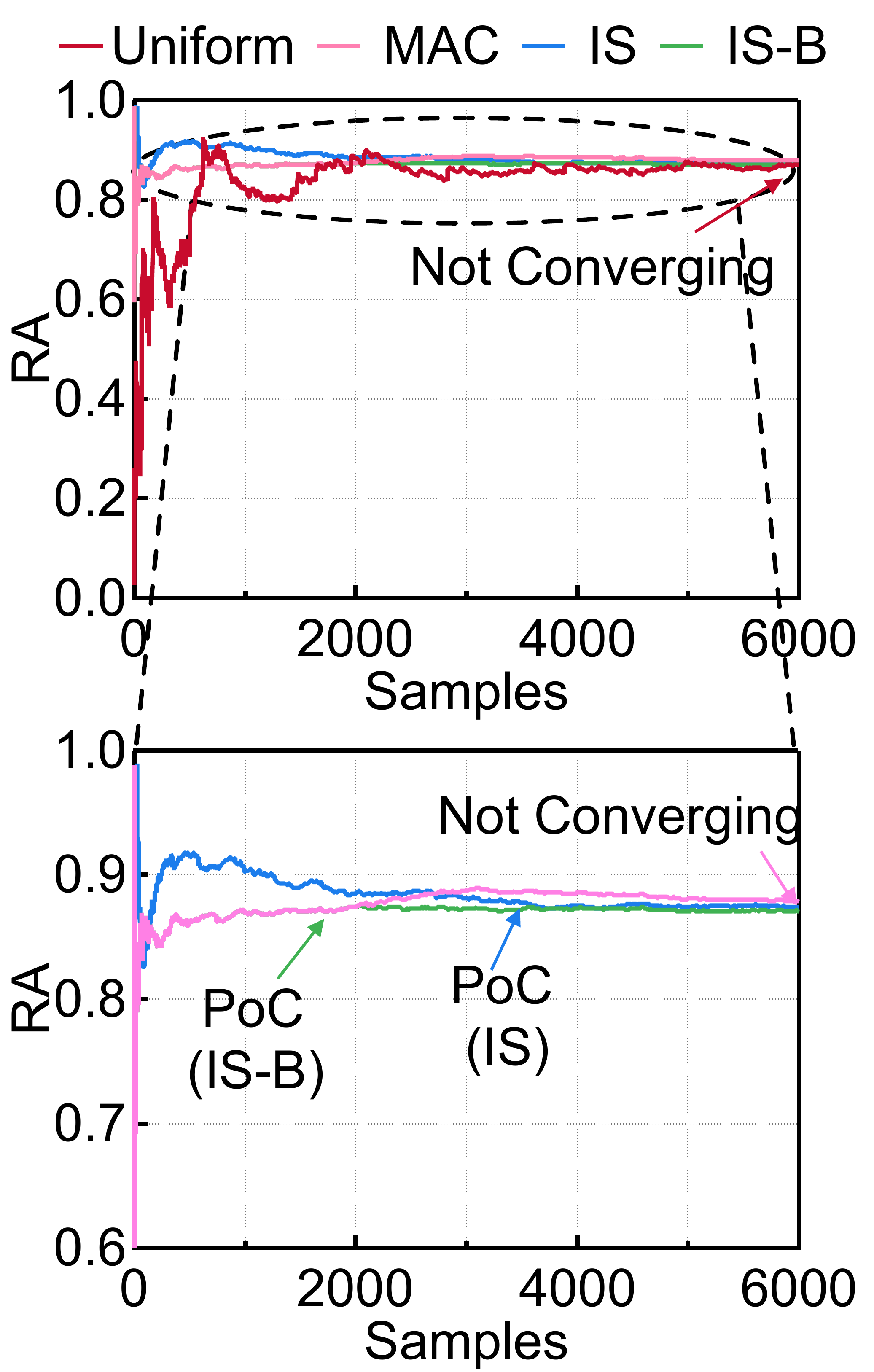}
  \label{fig:conv_alexnet}
}
\subfloat[\small{MNIST-Hogwild.}]
{
  \includegraphics[trim=0 0 0 0, clip, width=0.4\columnwidth]{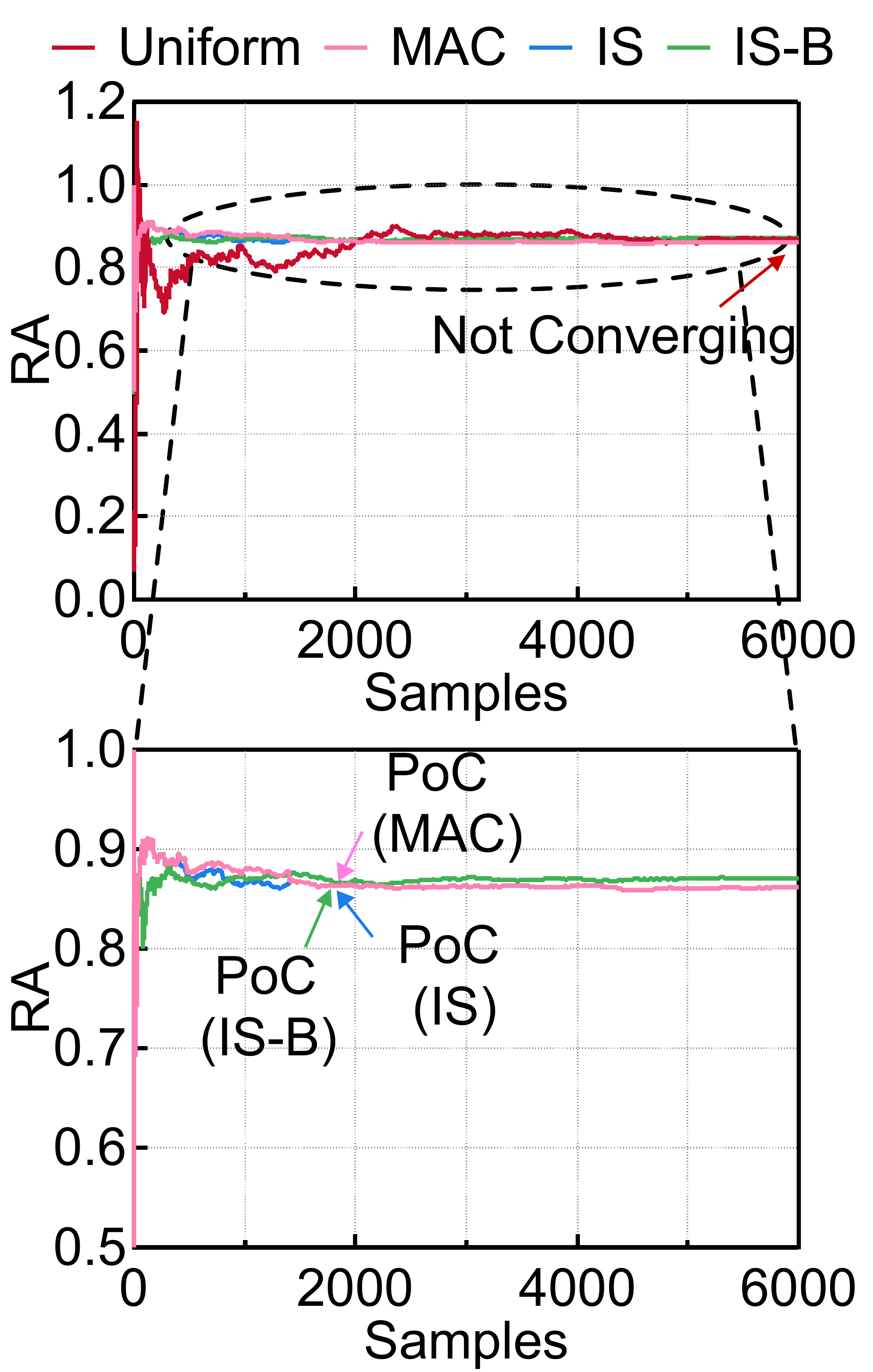}
  \label{fig:conv_alexnet}
}
\subfloat[\small{AlexNet.}]
{
  \includegraphics[trim=0 0 0 0, clip, width=0.4\columnwidth]{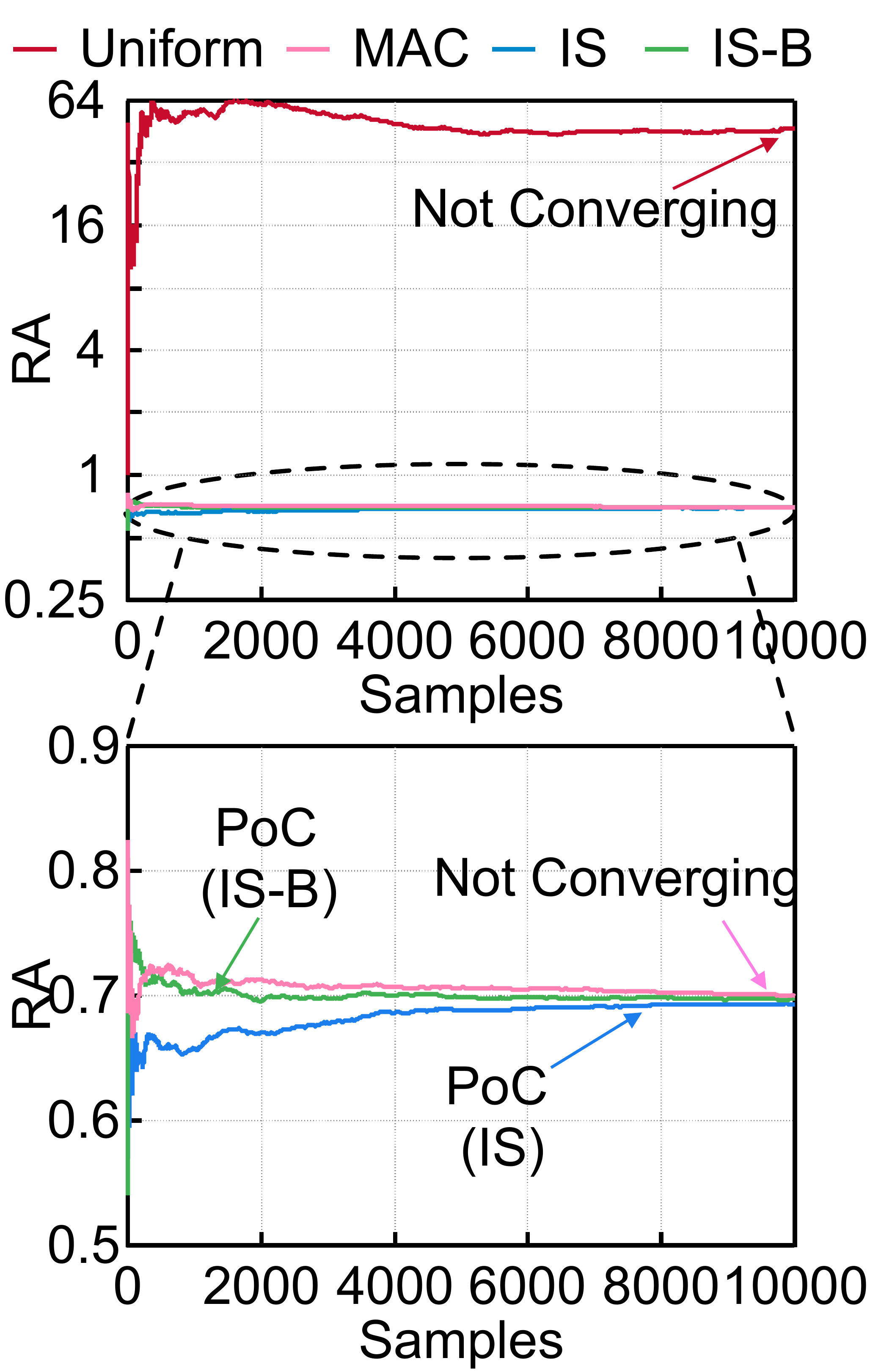}
  \label{fig:conv_alexnet}
}
\subfloat[\small{MobileNet.}]
{
  \includegraphics[trim=0 0 0 0, clip, width=0.4\columnwidth]{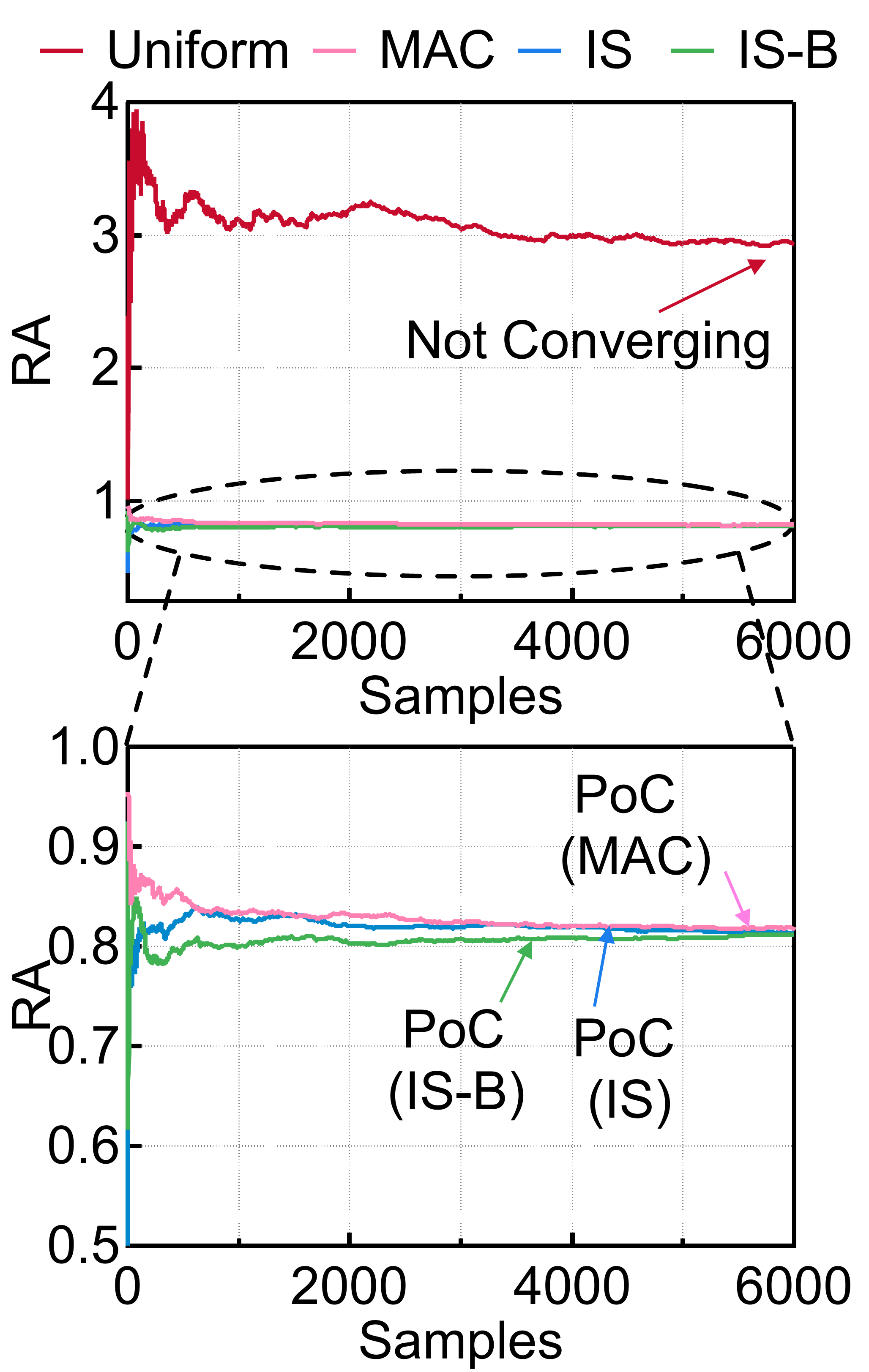}
  \label{fig:conv_mobilenet}
}
\subfloat[\small{YOLOv3.}]
{
  \includegraphics[trim=0 0 0 0, clip, width=0.40\columnwidth]{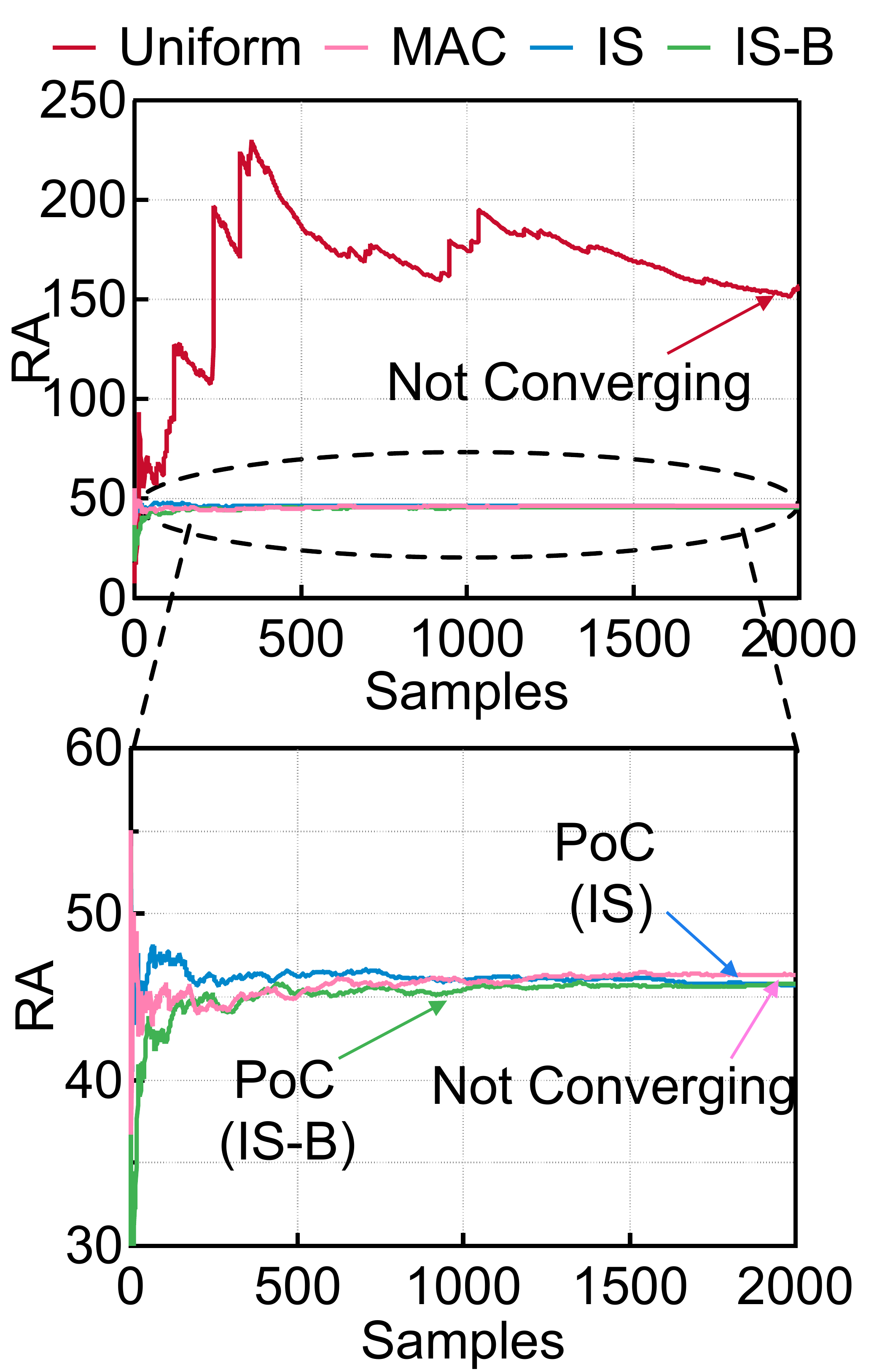}
  \label{fig:conv_yolo}
}
\caption{Convergence comparison of different sampling strategies. PoC stands for Point of Convergence. Overall, the order of convergence speed is \mode{IS-B} $>$ \mode{IS} $>$ \mode{MAC} $>$ \mode{Uniform}, which does not converge within given several thousands samples and produces physically unrealizable results.}
\vspace{-10pt}
\label{fig:convbignet}
\end{figure*}

We now evaluate the importance sampling strategy proposed in this paper on different DNNs and accelerators.

\begin{figure*}[t]
\centering
\subfloat[\small{AlexNet.}]
{
  \includegraphics[trim=0 0 0 0, clip, width=0.65\columnwidth]{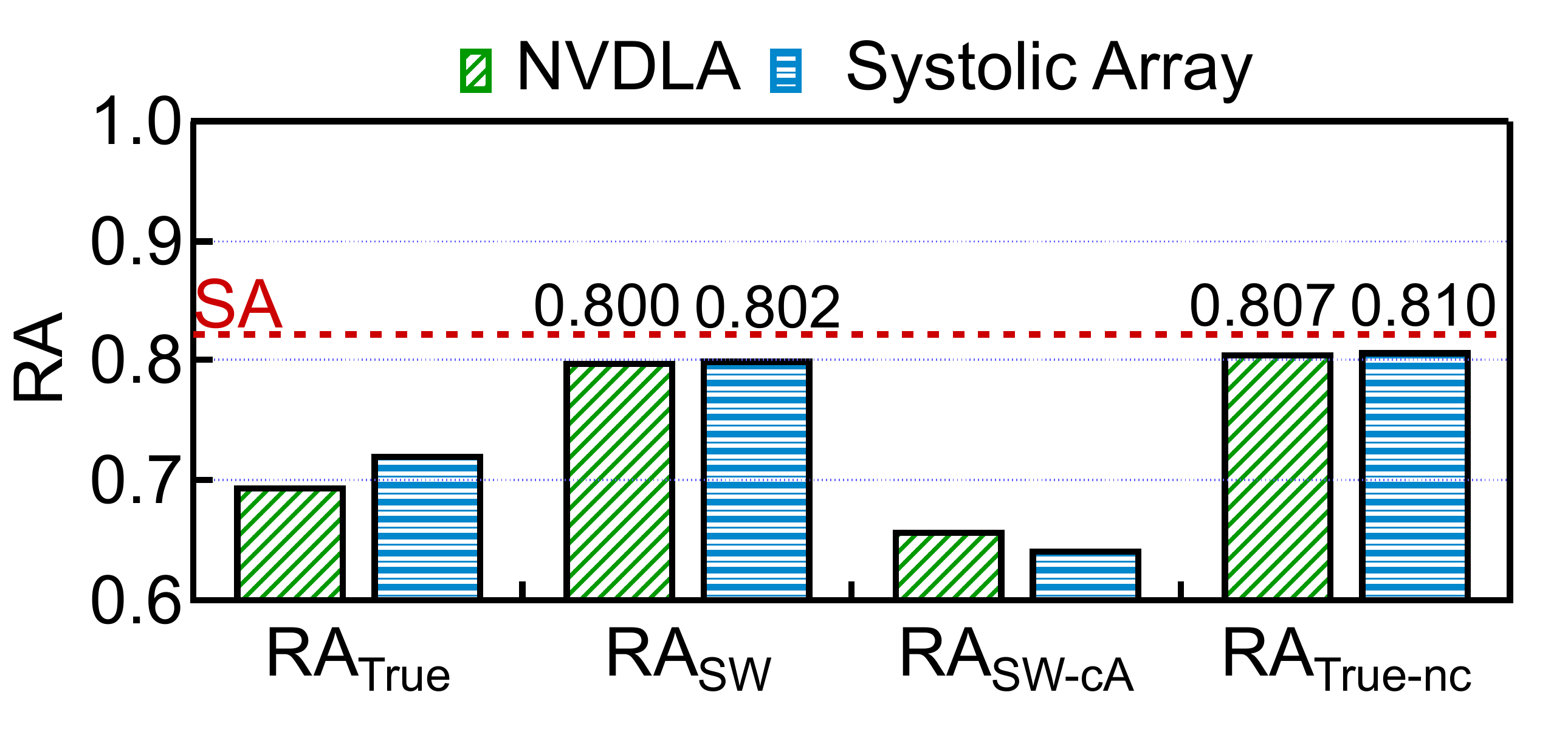}
  \label{fig:ra_alexnet}
}
\hfill
\subfloat[\small{MobileNet.}]
{
  \includegraphics[trim=0 0 0 0, clip, width=0.65\columnwidth]{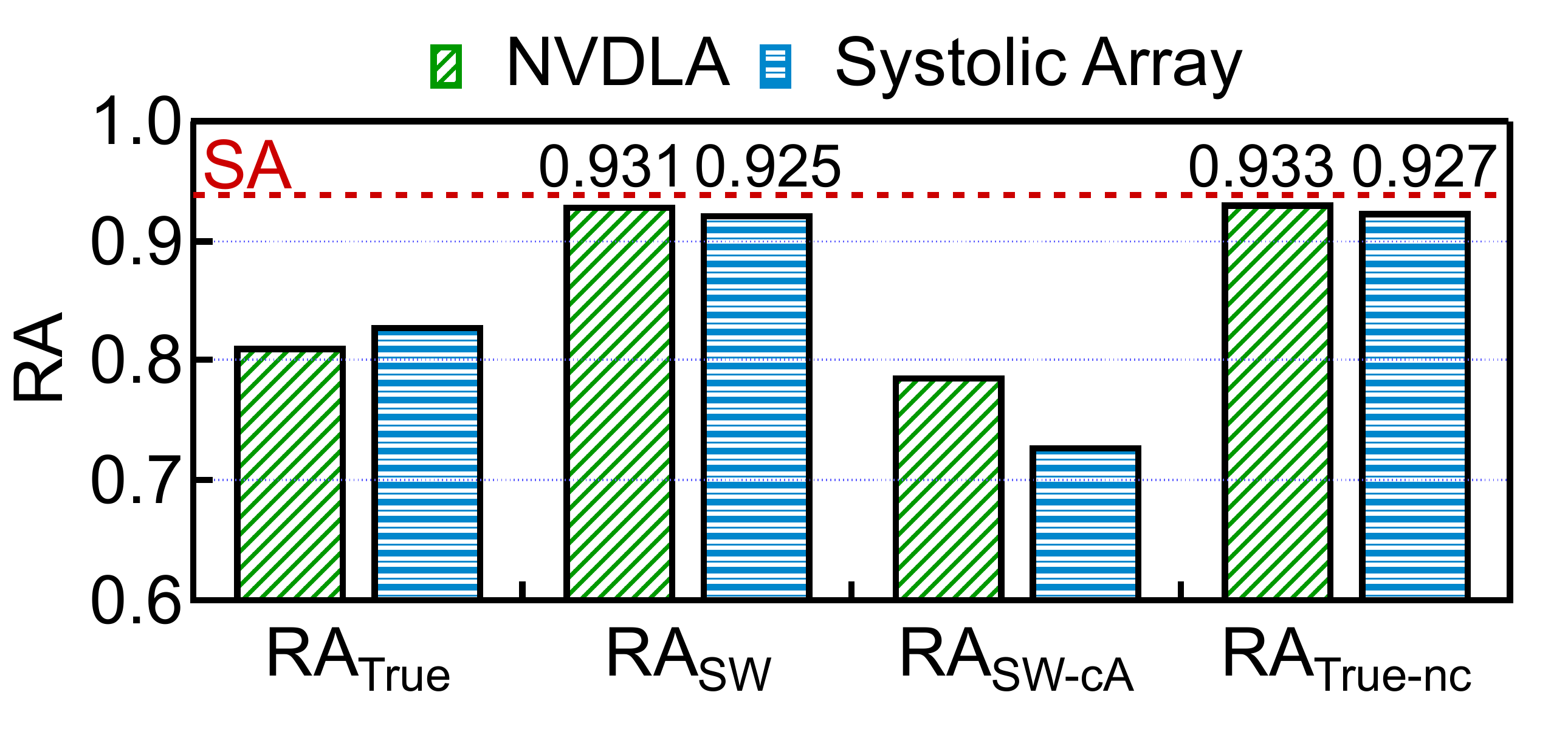}
  \label{fig:ra_mobilenet}
}
\subfloat[\small{YOLOv3.}]
{
  \includegraphics[trim=0 0 0 0, clip, width=0.65\columnwidth]{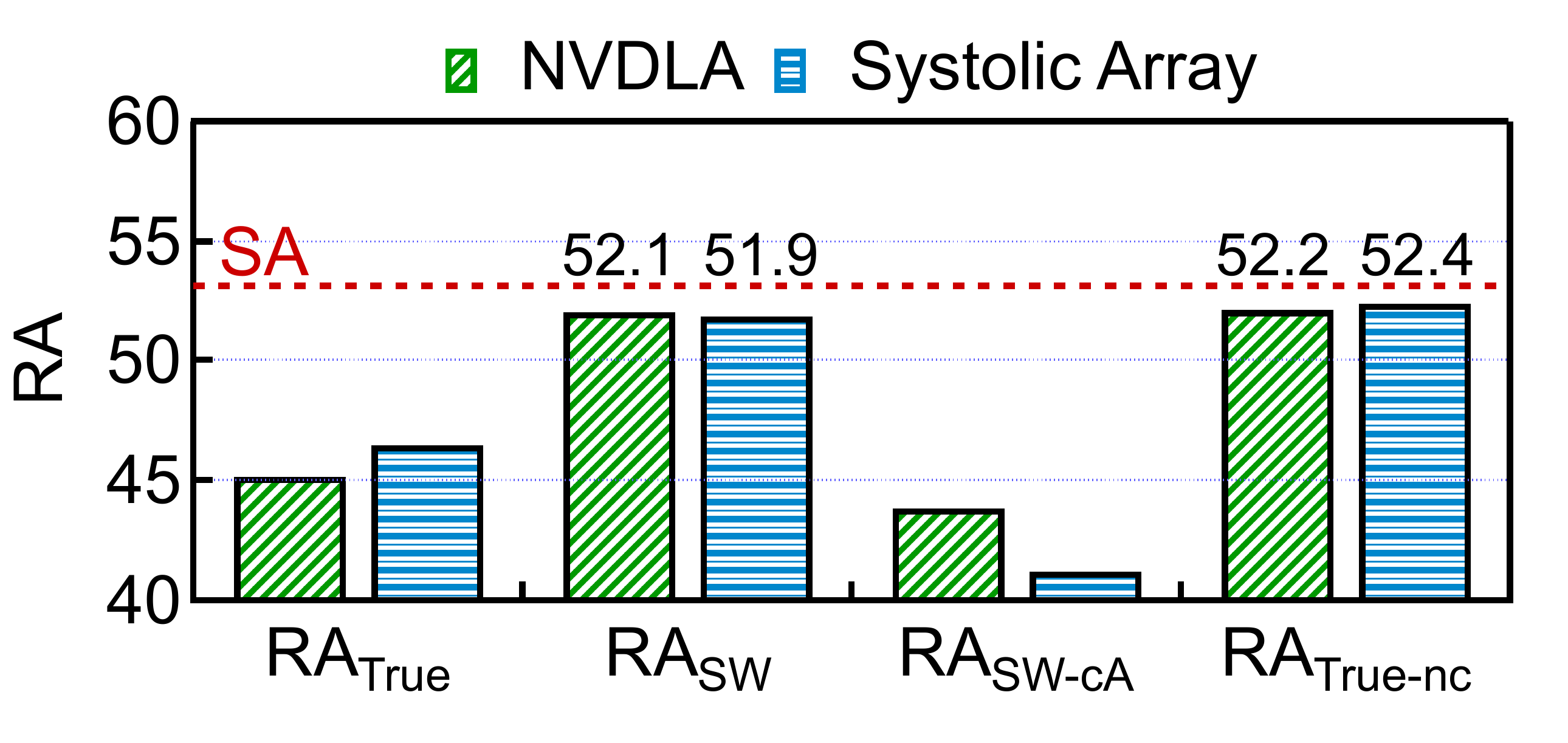}
  \label{fig:ra_yolo}
}
\caption{Comparison of RA formulations on the three large networks; the trend on the two small MNIST networks is similar. $RA_{True}$ is our RA formulation (\Equ{eq:rahw}); $RA_{Truc-nc}$ uses \Equ{eq:rahw} but assumes all control FFs are fault-free; $RA_{SW}$ is obtained using PyTorchFI and represents \Equ{eq:rasw}; $RA_{SW-cA}$ uses PyTorchFI but with the correct $A(j)$ estimation using FIdelity~\cite{he2020fidelity}.}
\vspace{-10pt}
\label{fig:rabignet}
\end{figure*}

\subsection{Experimental Setup}
\label{sec:ra:setup}

\paragraph{DNN Accelerators.} 
We evaluate on two DNN accelerators: a systolic array, whose configuration is the same as that used in \Sect{sec:val}, and Nvidia's NVDLA~\cite{nvdla}. We configure the NVDLA with 2 convolution cores, each with a $4\times4$ MAC units using the FP16 data type. Thse configuration is exactly the same as in FIdelity~\cite{he2020fidelity} , which also reports the FF distribution.


\paragraph{DNN Models.} We evaluate five common DNNs covering both image classification and object detection. Two DNNs are small in scale: a 5-layer LeNet~\cite{deng2012mnist} and a 4-layer MNIST-Hogwild~\cite{NEURIPS2019_9015}, on the MNIST dataset.
We also consider three larger networks: AlexNet~\cite{NIPS2012_c399862d} and MobileNet~\cite{howard2017mobilenets} for image classification on the CIFAR-10 dataset~\cite{krizhevsky2009learning}
and YOLOv3~\cite{redmon2018yolov3}, an object detection DNN, using the COCO dataset~\cite{lin2014microsoft}.

\paragraph{Evaluation Plan.} We aim to show that our sampling method 1) converges to the ground truth RA, and 2) converges faster than existing sampling methods.

To obtain the ground-truth RA, we exhaust all the fault sites for two small networks LeNet (352K) and MNIST-HogWild (412K), but for three larger networks we evaluate, it would take unrealistically long time. Instead, our strategy is to intentionally target a subset of the hardware fault sites and validate within that subset. Specifically, we randomly sample 500K fault sites for AlexNet, MobileNet, and YOLOv3., and assume that these are all the hardware fault sites. All subsequent evaluations are restricted to only those fault sites. The raw FIT rate is the same as described in \Sect{sec:val:prob}


\subsection{Comparing Sampling Methods}
\label{sec:ra:sample}

We also aim to compare how effective different sampling strategies are for estimating $RA_{True}$. To that end, we will compare the following four sampling methods:

\begin{itemize}
  \item \mode{Uniform}: this strategy draws a software fault site uniformly at random.

  \item \mode{MAC}: this sampling strategy is adapted from Mahmoud et al.~\cite{mahmoud2021optimizing}; the probability of a weight or an activation getting sampled in fault injection is proportional to the number of MAC operations they are involved in.

  \item \mode{IS}: this is importance sampling using the constancy heuristic in \Sect{sec:sampling:is}, where the PDF for sampling is proportional to the faulty probabilities of the software fault sites $p(j)$ while assuming $A(j)$ is constant.

  \item \mode{IS-B}: this is importance sampling using the BP heuristic in \Sect{sec:sampling:is}. $A(j)$ that considers the different accuracy impacts of different BPs. We randomly inject faults into different BPs and observe the accuracy drop.
  Averaging our profiling results, we assume an inference accuracy drop of 15\% for the MSB in the exponent and an 8\% accuracy drop for the next four exponent bits; the accuracy drop of other BPs are assumed to be zero.

\end{itemize}


The convergence curves of different sampling methods are shown in \Fig{fig:convbignet}.
The Point of Convergence (PoC) must satisfy two criteria: the average RA in a window of 300 samples must be within 0.3\% of the ground truth and the variance in the same window must be below $10^{-2}$.

Immediately noticeable is that the RA under uniform sampling for the three large networks is well above one, which is physically impossible, indicating that \mode{Uniform} would take a lot more samples to converge.
The reason that \mode{Uniform} has a hard time converging is that the probabilities of the software fault sites vary by orders of magnitude, for which uniform sampling is inefficient (\Sect{sec:sampling}). For instance, in YOLOv3 the faulty probability of a weight of the first convolution layer is two orders of magnitude higher than that of in the twelfth convolution layer $1.61\times10^{-4}$\% vs. $2.67\times10^{-6}$\%).
The same difference exists within a layer, too.

Comparing other sampling strategies, we observe a general trend, \mode{IS-B} converges faster than \mode{IS}, which is faster than \mode{MAC}, confirming the benefits of using a sampling PDF that is more proportional to the integrand during Monte Carlo integration.
\mode{MAC} converges more slowly than the two importance sampling-based methods, because \mode{MAC}, while considers different reuse factors of weights/activations based on the MAC operations, does not take into account the actual FF distribution in the underlying hardware architecture.





%% file: compare.tex
\section{Comparative Studies}
\label{sec:compare}

\paragraph{Comparing RA Methods.}
Using the three larger networks and the setup described in \Sect{sec:ra:setup}, \Fig{fig:rabignet} compares the RA estimated from a set of different formulations/methods:

\begin{itemize}
    \item $RA_{True}$: RA estimated using our formulation.
    \item $RA_{True-nc}$: RA estimated using our formulation, but assuming all global FFs are fault free.
    \item $RA_{SW}$: RA estimated using the formulation in \Equ{eq:rasw}, which represented the method used by common tools today such as PyTorchFI and TensorFI.
    \item $RA_{SW-cA}$: RA estimated using the \Equ{eq:rasw} formulation but using the correct $A(j)$ estimated from FIdelity~\cite{he2020fidelity}, but still (incorrectly) assumes that all software fault sites are equally faulty.
\end{itemize}

\no{As a reference, we show the SA for each network.
Not surprisingly, all DNNs have lower RAs, regardless of the formulation, than their corresponding SAs, indicating significant accuracy drops under transient errors.}

$RA_{SW}$ is always higher than $RA_{True}$, indicating an over-estimation. In fact, under $RA_{SW}$ a DNN's RA is very close to SA. For instance, AlexNet has only a 0.7\% accuracy drop and YOLOv3 has only a 1.2 mAP drop compared to their SAs, giving a perhaps rather optimistic impression that today's DNNs are resilient to transient faults.

A primary source of over-estimation in $RA_{SW}$ is that software fault injection does not consider global control FFs, which in NVDLA contribute to 11.3\% of the total FFs.
To tease apart the effects of global control FFs and other FFs, we compare $RA_{True-nc}$ with $RA_{True}$. ``$RA_{True}$'' is over 10\% lower than ``$RA_{True-nc}$'', matching our intuition that application crashes dramatically lower the accuracy.

$RA_{True-nc}$ is slightly, but consistently, higher than $RA_{SW}$ across all DNNs even though both do not consider control FFs. This is because $RA_{SW}$ assumes that a corrupted weight affects all the output activations, where in reality the affected neurons are fewer due to the limited reuse factor of a FF.

$RA_{SW-cA}$ consistently under-estimates the RA compared to $RA_{True}$.
The reason is that a uniform faulty probability distribution across software fault sites made by $RA_{SW-cA}$ exaggerates the accuracy degradation effect caused by the control FFs, which occupy a small portion of the total FFs but result in much lower (zero) RAs given bit flips. 


\begin{figure}[t]
\centering
\includegraphics[width=0.8\columnwidth]{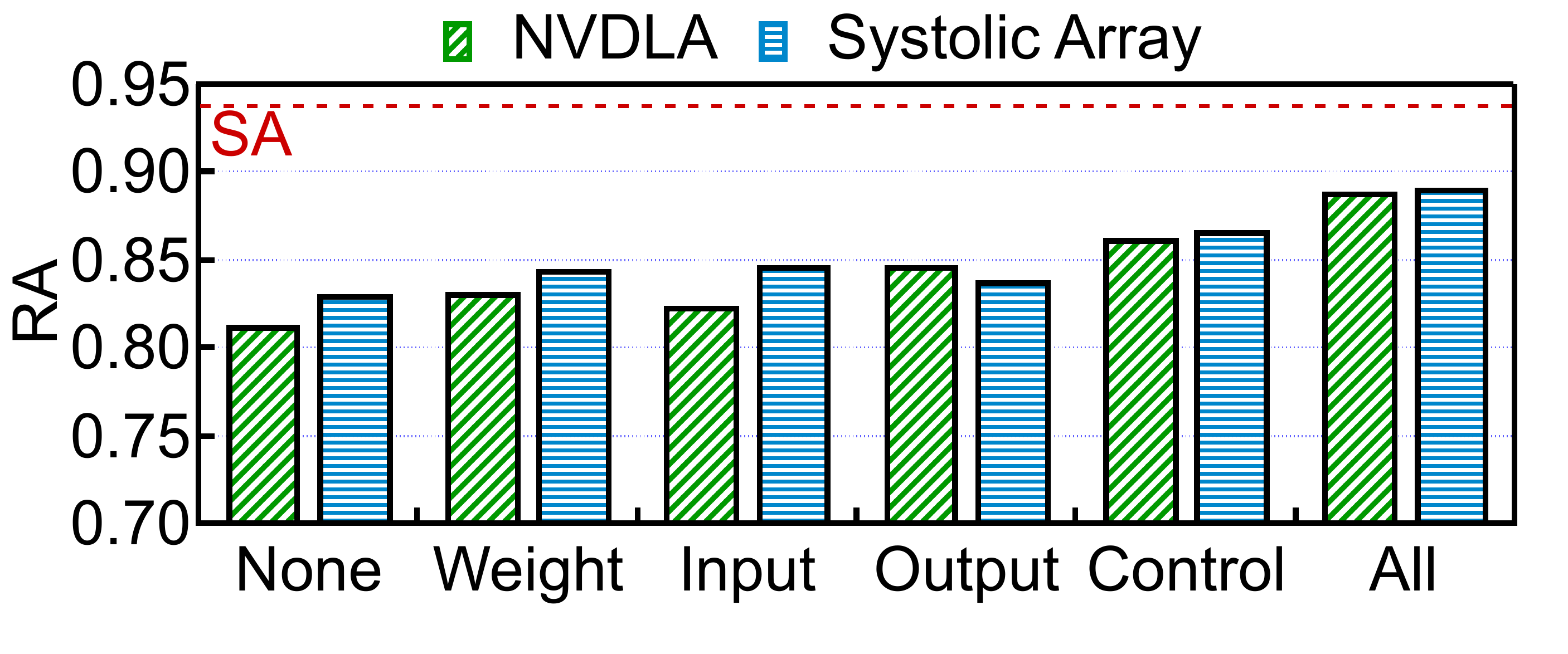}
\caption{We pick one FF type ($x$-axis) to harden at a time and evaluate the resulting MobileNet RA ($y$-axis).}
\vspace{-10pt}
\label{fig:fit}
\end{figure}

\paragraph{Comparing Networks.} 
YOLOv3 has a lower SA to RA drop compared to the two image classification networks AlexNet and MobileNet. Specifically, AlexNet and MobileNet have a SA-to-RA drop of 11.4\% and 12.0\%, whereas the SA-to-RA drop for YOLOv3 is 8.2\%. The reason is that object detection is generally more resilient: a slight change in bounding box position and size can be tolerated as long as its intersection with the ground truth bounding box over the union of them is higher than the threshold.


\paragraph{Comparing Accelerators.} We make two observations when comparing the results between NVDLA and systolic array.
First, $RA_{True}$ on NVDLA is higher than that on the systolic array. This is due to the absence of the FFs before the on-chip memory in our simple systolic array design. Faults in FFs before the on-chip memory corrupt data in the SRAM and propagate faults to a significantly larger number of neurons. 
Second, $RA_{SW-cA}$ on NVDLA is higher than that on the systolic array. This difference, again, arises from the absence of FFs before the on-chip memory in the systolic array. Without those FFs, the faulty probability for control FFs increases, reducing $RA_{SW-cA}$ on the systolic array.

\paragraph{Sensitivity of Raw FIT Rates.}
Hardening FFs in hardware is a widely used technique to mitigate transient errors~\cite{rajaei2015single, peng2018radiation}. To study the impact of FF hardening on RA, we perform an experiment where we harden one FF type at a time. We assume that FFs are hardened using the common Dual Interlocked storage Cell (DICE) technology~\cite{calin1996upset}, and the raw FIT rate for DICE-hardened FFs is 200/MB (cf. 600/MB for unhardened FFs) following Jagannathan et al.~\cite{jagannathan2012frequency}. 

\Fig{fig:fit} shows the resulting RA of MobileNet ($y$-axis); the $x$-axis shows the FF type that is hardened. ``None'' and ``All'' mean no and all FF types are hardened, respectively.
Unsurprisingly, RA with one hardened FF type is between that of ``None''and ``All''.

Our results provide insight into which category of FFs should be hardened for improving RA. For instance, hardened the control FFs has the highest RA improvements, since transient faults on control FFs usually crash an accelerator altogether.
Interestingly, while in NVDLA output FFs are the largest in number of all categories, hardening them does not provide the highest RA improvement. This is because the reuse factor of output FFs is the smallest (one), so transient faults on them have low impact on accuracy.

\begin{figure}[t]
\centering
\subfloat[\small{20\% threshold.}]
{
  \includegraphics[trim=0 0 0 0, clip, width=0.48\columnwidth]{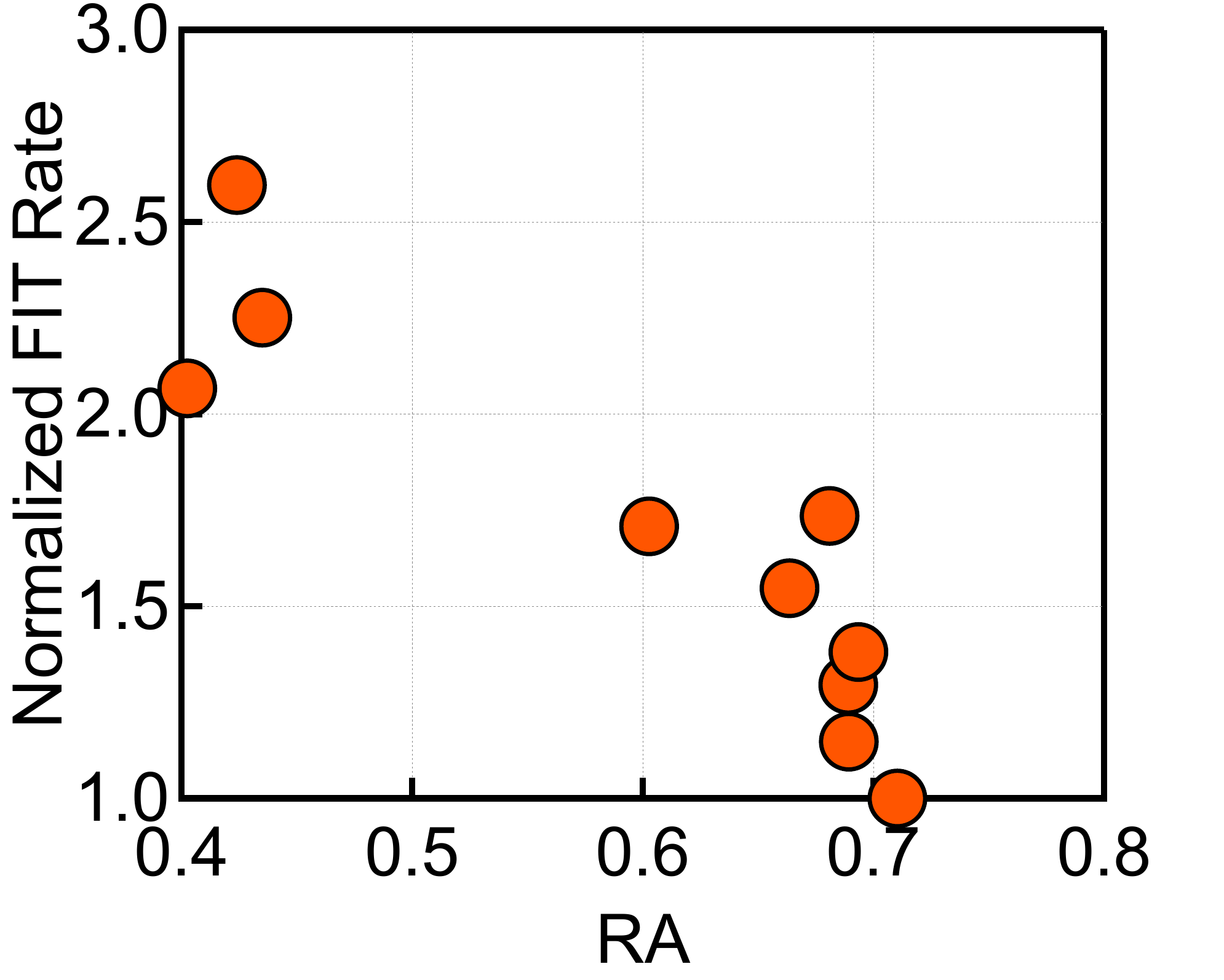}
  \label{fig:rafit20}
}
\subfloat[\small{40\% threshold.}]
{
  \includegraphics[trim=0 0 0 0, clip, width=0.48\columnwidth]{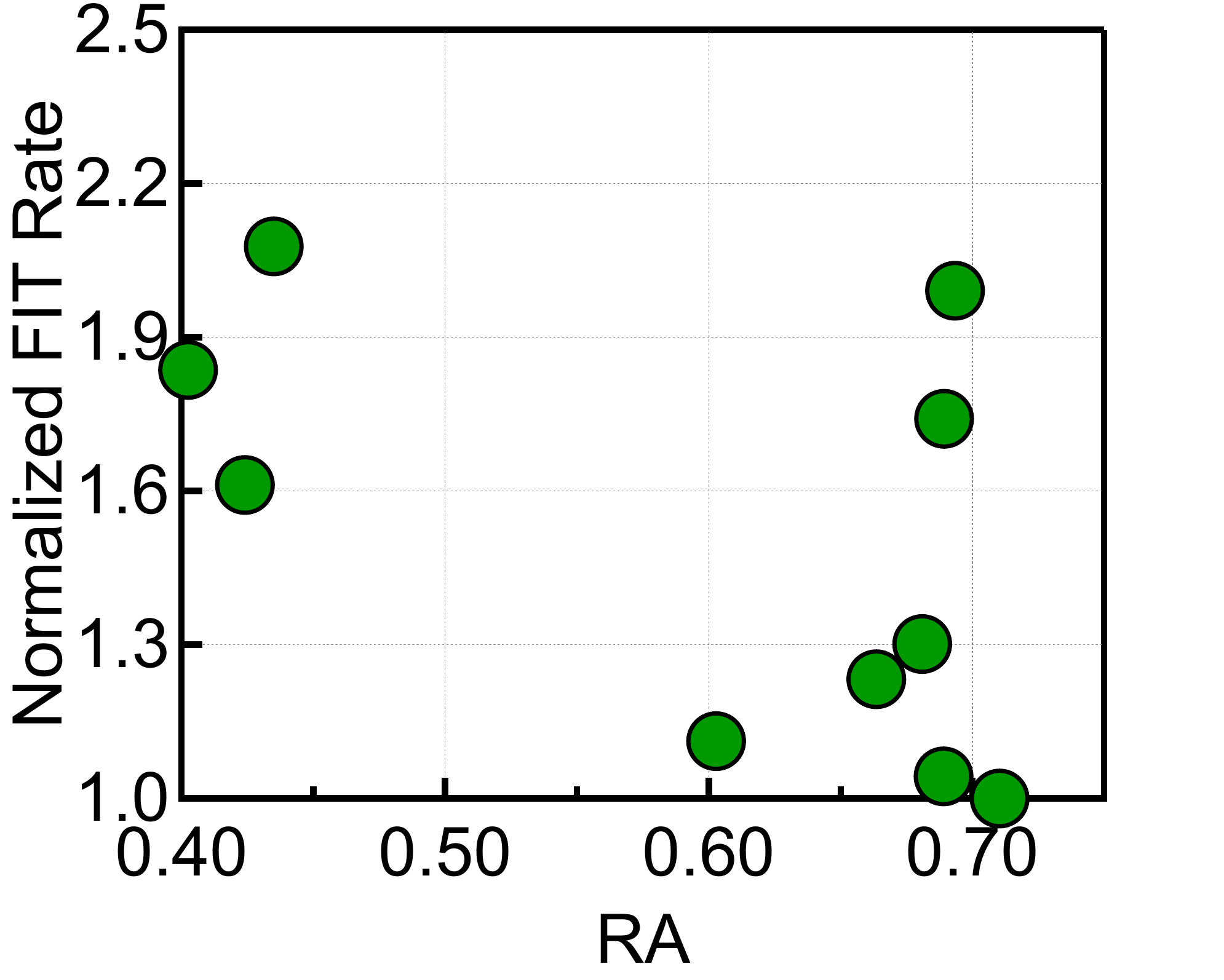}
  \label{fig:rafit40}
}
\caption{RA ($x$-axis) vs FIT rate ($y$-axis) correlation under two thresholds for ten NAS-generated networks.}
\vspace{-10pt}
\label{fig:ra_fit}
\end{figure}

\mbox{\paragraph{RA, FIT, and SDC Correlation.}}
We use ten networks found by the NAS framework (\mbox{\Sect{sec:nas}}) to study the RA vs FIT rate correlation. Like prior work on DNN resiliency~\mbox{\cite{li2017understanding, he2020fidelity}}, we use a threshold $T$ in estimating the FIT rate. Specifically, a transient fault is considered to generate a failure if the inference accuracy under that fault is lower than the fault-free accuracy by $T$. Similar to Fidelity~\mbox{\cite{he2020fidelity}}, we use two thresholds for this study: 20\% and 40\%.


\mbox{\Fig{fig:ra_fit}} shows the results under the two thresholds. The FIT rate is normalized to the lowest FIT rate among the ten networks. While in general RA is inversely correlated with the FIT rate, notable outliers exist especially when the threshold is large: the Pearson correlation coefficients are -0.90 and -0.53 under the 20\% and 40\% threshold, respectively. In addition, the FIT rate does not provide the actual inference accuracy results, which could be critical to algorithm designers.

SDC rate is similarly defined as the FIT rate (and thus shares the same caveats above) except SDC does not consider the potential system crashes under transient errors, which is non-trivial. On CifarNet~\mbox{\cite{hosang2015taking}}, ResNet-18~\mbox{\cite{he2016deep}}, DETR-R50~\mbox{\cite{carion2020end}}, and DeepSpeech~\mbox{\cite{hannun2014deep}}, 6.86\%, 7.13\%, 5.55\% and 4.79\% of the transient faults yield crashes, respectively.


%% file: nas.tex
\section{RA-Aware NAS}
\label{sec:nas}

We show how our RA estimation can help automatically design DNNs with high RAs through Network Architecture Search (NAS). We first introduce our RA-aware NAS framework (\Sect{sec:nas:algo}) followed by the experimental setup (\Sect{sec:nas:exp}). We show that networks generated using our NAS framework possess significantly higher RA than NAS that uses other RA estimation methods (\Sect{sec:nas:res}).

\subsection{The NAS Framework}
\label{sec:nas:algo}

As a case study, we focus on multi-modal DNNs, which fuse two or more data modalities and are widely used in application domains such as autonomous machines~\cite{kato2018autoware, ap}.
We study multi-modal fusion DNNs for two reasons. First, they are widely used in applications such as autonomous machines, where resiliency against soft errors is critical. For instance, autoware~\cite{kato2018autoware} and Baidu Applo~\cite{ap} fuse RGB cameras and LiDAR-generated point clouds for object tracking. Second, these networks are challenging workloads for NAS, because they expose additional degree of freedom such as at what layers and using what operators to fuse different modalities.

We propose a RA-aware NAS algorithm for search fusion DNNs.
Our NAS algorithm is built on top of the MFAS fusion NAS framework~\cite{perez2019mfas}, which was originally designed to search fusion networks with the highest standard accuracy. We claim \textit{no} algorithmic novelty over MFAS. Rather, our intention is to show that our RA estimation algorithm can be readily integrated into existing NAS frameworks, such as MFAS, to enable RA-aware NAS.

\paragraph{Problem Setup.}
We consider a basic fusion DNN containing two branches, each processing a particular input modality (e.g., RGB images vs. point clouds).
The intermediate feature maps at certain layers of each branch are fused before going through a unified branch to generate the output.
The searching space includes three degrees of freedom:

\begin{enumerate}
    \item the number of layers, $L$, from each branch to fuse; note that each branch could have more than $L$ layers, but only $L$ layers from each branch are used for fusion;
    \item the exact $L$ layers from the first branch ($X_1, ..., X_L$) and the exact $L$ layers from the second branch ($Y_1, ..., Y_L$) to fuse; layers are fused in order --- that is, $X_i$ and $Y_i$ are implicitly a pair of fusion layers;
    \item the activation function after each of the $L$ fusion pairs.
\end{enumerate}

Our goal is to identify a network whose fusion configuration maximizes the RA.
\no{The search space is vast. Considering only 5 layers in each branch, there are over 33 million configurations, which would take over 1,000 centuries to search on two Nvidia 2080 Ti GPUs in our measurement.}

\no{\paragraph{RA-aware Search Algorithm.}
To tame the vast search space, we follow a common idea in the NAS literature, i.e., training a lightweight surrogate function that predicts the RA given a fusion configuration~\cite{liu2018progressive,perez2018efficient}.}

\no{In particular, we first enumerate all $K_1$ configurations assuming only 1 fusion pair; we train the $K_1$ networks to convergence. For each of the $K_1$ networks, we obtain its RA using our importance sampling-based algorithm described in \Sect{sec:sampling}.
These $K_1$ networks are then used to train a surrogate function, a one layer LSTM, which predicts RA from a particular fusion configuration. The surrogate function is used to predict the RAs of networks with 2 fusion pairs. From the predictions, we obtain the top $K_2$ configurations with highest RAs. These $K_2$ and their corresponding actual RAs are used to fine-tune the surrogate function, which is then used to predict RAs of networks that have 3 fusion pairs.

This process continues until we reach $L$ fusion pairs, at which point the surrogate will have predicted the RAs of networks that have $L$ fusion pairs, from which we pick the top $K_5$ networks and obtain their actual RAs. The one with the highest RA is the NAS output.}




\subsection{Experimental Setup}
\label{sec:nas:exp}

\paragraph{Network and Datasets.} We target two-branch fusion networks for action recognition. The first branch takes RGB videos as the input and the second branch takes skeleton-based pose data as the input. Following the setup in MFAS~\cite{perez2019mfas}, we use the inflated ResNet-50~\cite{baradel2018glimpse} for the video branch and the deep co-occurrence model~\cite{li2018co} for the pose branch. We use the NTU RGB+D dataset~\cite{shahroudy2016ntu}. We set the maximum $L$ to be 3, i.e., at most three fusion layers are allowed in NAS.

\paragraph{Evaluation Plan.} We have two aims. First, we aim to show that our RA estimation algorithm allows the NAS framework to identify networks with higher RAs than today's software-level RA estimation algorithms. Second, we aim to show that our importance sampling method improves the NAS speed over random sampling.



\subsection{Analyzing the Networks}
\label{sec:nas:res}

\begin{figure}[t]
\centering
\begin{minipage}[t]{0.48\columnwidth}
  \centering
  \includegraphics[width=\columnwidth]{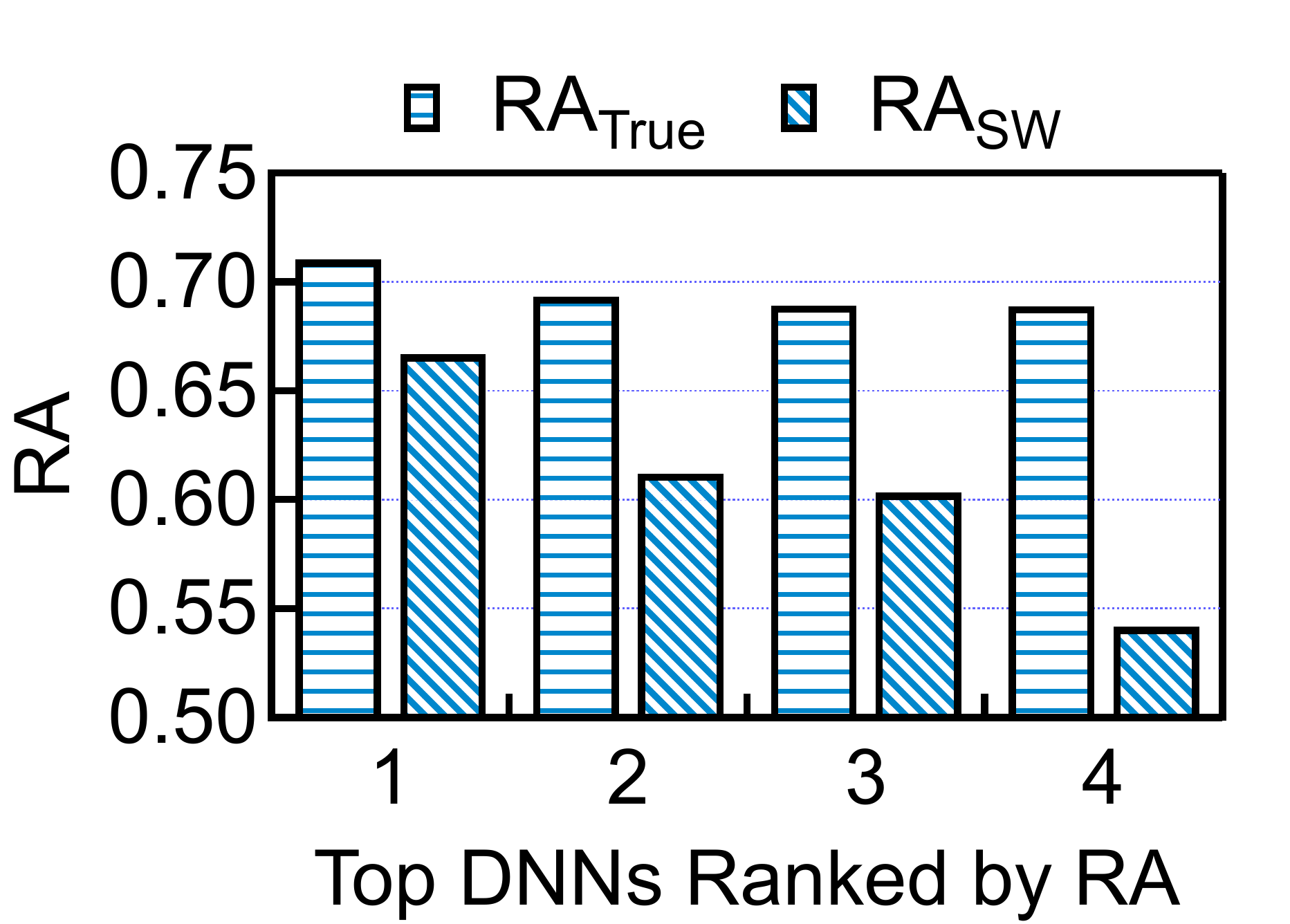}
  \caption{Top architectures' RA comparison between using $RA_{True}$ and $RA_{SW}$.}
  \label{fig:racomp}
\end{minipage}
\hspace{2pt}
\begin{minipage}[t]{0.48\columnwidth}
  \centering
  \includegraphics[width=\columnwidth]{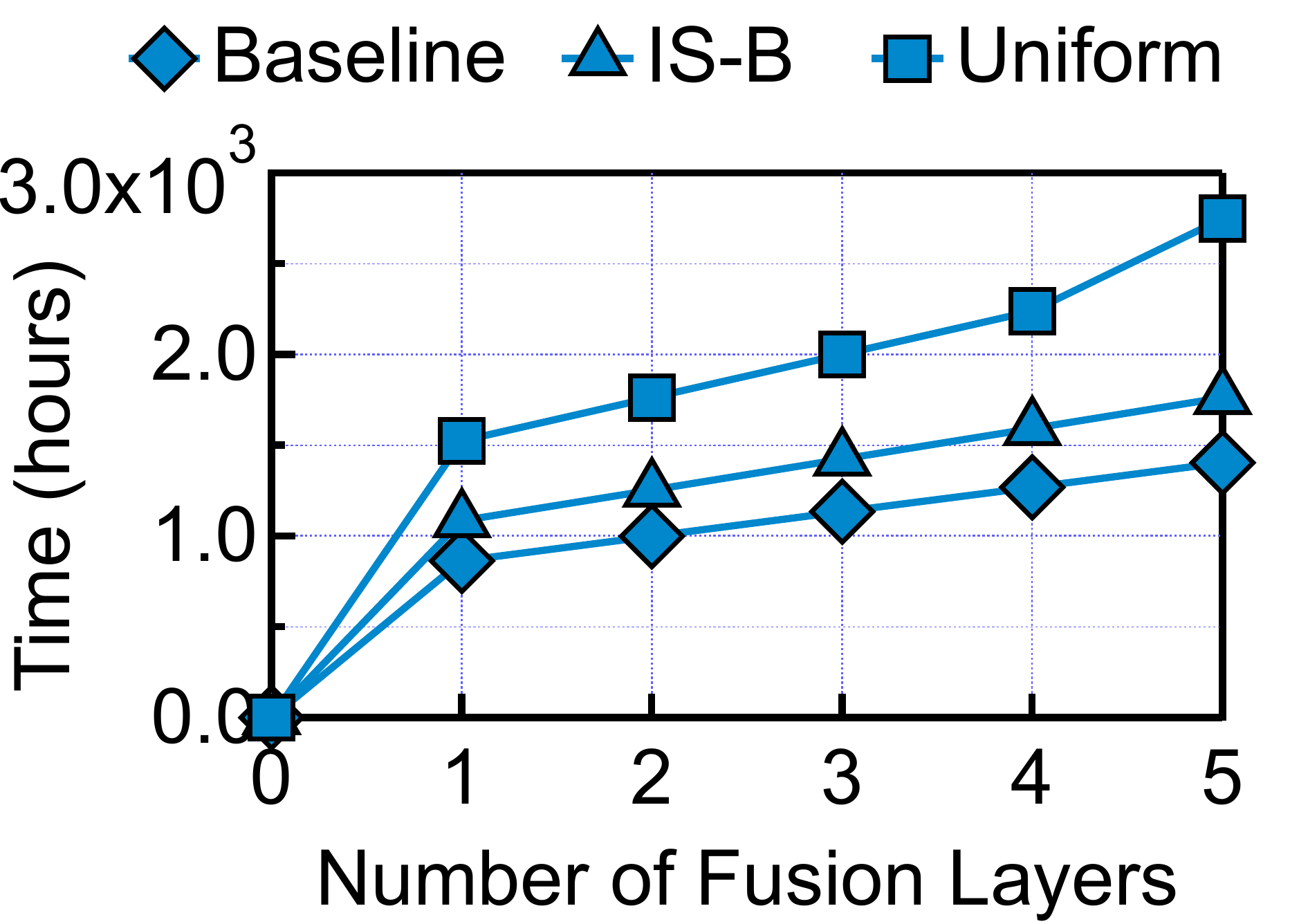}
  \caption{NAS latency of using \mode{IS-B} and \mode{Uniform}.}
  \label{fig:latcomp}
\end{minipage}
\vspace{-5pt}
\end{figure}

We show that NAS guided by our RA estimation algorithm $RA_{True}$ yields networks that have higher RAs than those guided by $RA_{SW}$.
\Fig{fig:racomp} compares the RAs of the top 4 networks generated by NAS using the two RA estimation methods. Note that the RAs of all networks, regardless of how they are generated, are all evaluated using our RA estimation algorithm to obtain the true RAs. 
The average improvement of RA is 9.0\% across the four networks. In fact, the fourth resilient network using $RA_{True}$ to guide NAS has a higher RA compared to the best network searched using $RA_{SW}$.

\no{\Fig{fig:nasresult} shows the highest-RA networks generated under two RA methods, $RA_{True}$ (\Fig{fig:nasresult}) and $RA_{SW}$ (\Fig{fig:nasrasw}), which have an RA of 71.0\% and 66.7\%, respectively.}

\begin{table}[t]
\centering
\caption{Top five networks generated by NAS when the number of fusion layers $L$ is 2 vs. 3 ($L$ is capped at 3). Each <$X, Y$> denotes a fusion pair. For instance, <4, 1> means the output of layer 4 in the first branch and the output of layer 2 in the second branch are fused together. }
\renewcommand*{\arraystretch}{1}
\renewcommand*{\tabcolsep}{8pt}
\resizebox{.95\columnwidth}{!}
{
\begin{tabular}{cc|cc}
\toprule[0.15em]
\textbf{RA (L=2)} & \textbf{<X,Y>}  & \textbf{RA (L=3)} & \textbf{<X,Y>}  \\
\midrule[0.05em]
40.2\%       & <4,2>,<2,2>  & 66.4\%  & <4,3>,<3,2>,<3,1>\\ 
42.4\%         & <4,3>,<3,4>  & 68.1\% & <2,1>,<4,4>,<4,2>\\ 
43.5\%        & <4,3>,<2,3>  & 68.9\%  & <4,3>,<3,2>,<4,1>\\
60.3\%       & <2,1>,<4,4> & 69.3\%  & <4,3>,<2,3>,<4,1>\\
68.9\%         & <4,3>,<3,2>  & 71.0\%  & <4,3>,<3,4>,<4,1>\\
\bottomrule[0.15em]
\end{tabular}
}
\vspace{-3mm}
\label{tbl:nas}
\end{table}

\paragraph{Analyzing the NAS-ed Networks.} 
We find that fusion DNNs that have higher RAs share two common characteristics.

First, using more fusion layers generally improve the resilience of the fusion networks. \Tbl{tbl:nas} compares the RAs of the top five DNNs when $L=2$ (two fusion layers) vs. $L=3$ (three fusion layers). By adding one more fusion layer, NAS is able to improves the highest RA of a network by 2.1\% and the average RA of the top five networks by 17.8\%.
\no{This also explains why the network in \Fig{fig:nasrasw}, generated when using $RA_{SW}$ to guide NAS, has a lower RA --- it has only two rather than three fusion layers.}

Second, high-RA DNNs usually fuse feature maps from later stages of both branches. \Tbl{tbl:nas} shows the fusion configuration of the top-RA DNNs, where <$X, Y$> indicates that layer $X$ from the first branch and layer $Y$ from the second branch are fused. All these networks make extensive use of the last two layers for fusion (i.e., frequent use of <$4, 3$>, <$3, 4$>, and <$4, 4$>). That said, while earlier layers are rarely fused with each other, they do participate in certain fusion pairs, indicating the need to mix different layers in fusion.


\no{
\begin{figure}[t]
\centering
\subfloat[Network with the highest RA generated using $RA_{True}$.]{
    \label{fig:nasresult}
    \includegraphics[height=1.11in]{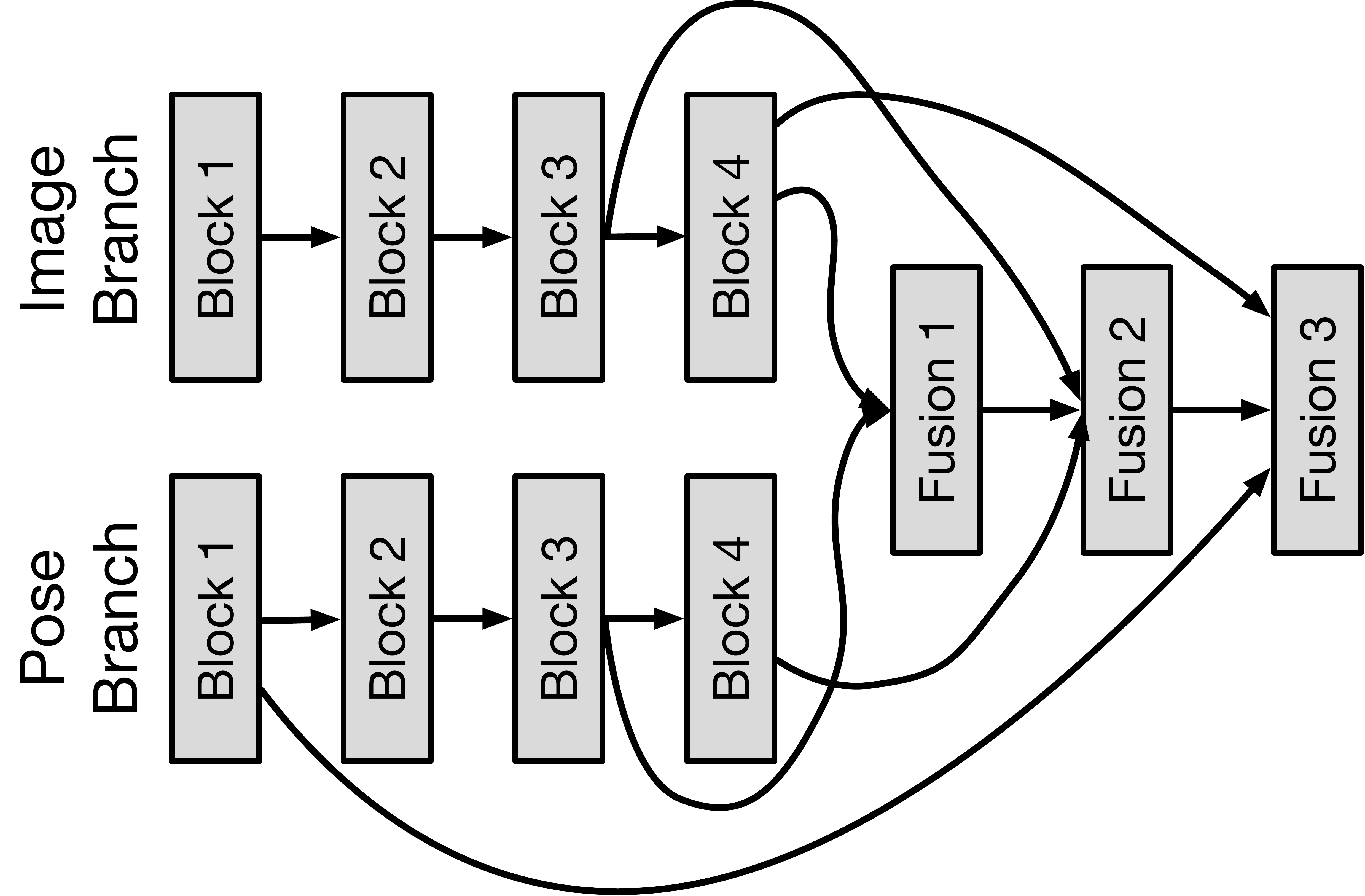}
}
\hfill
\subfloat[Network with the highest RA generated using $RA_{SW}$.]{
    \label{fig:nasrasw}
    \includegraphics[height=1.11in]{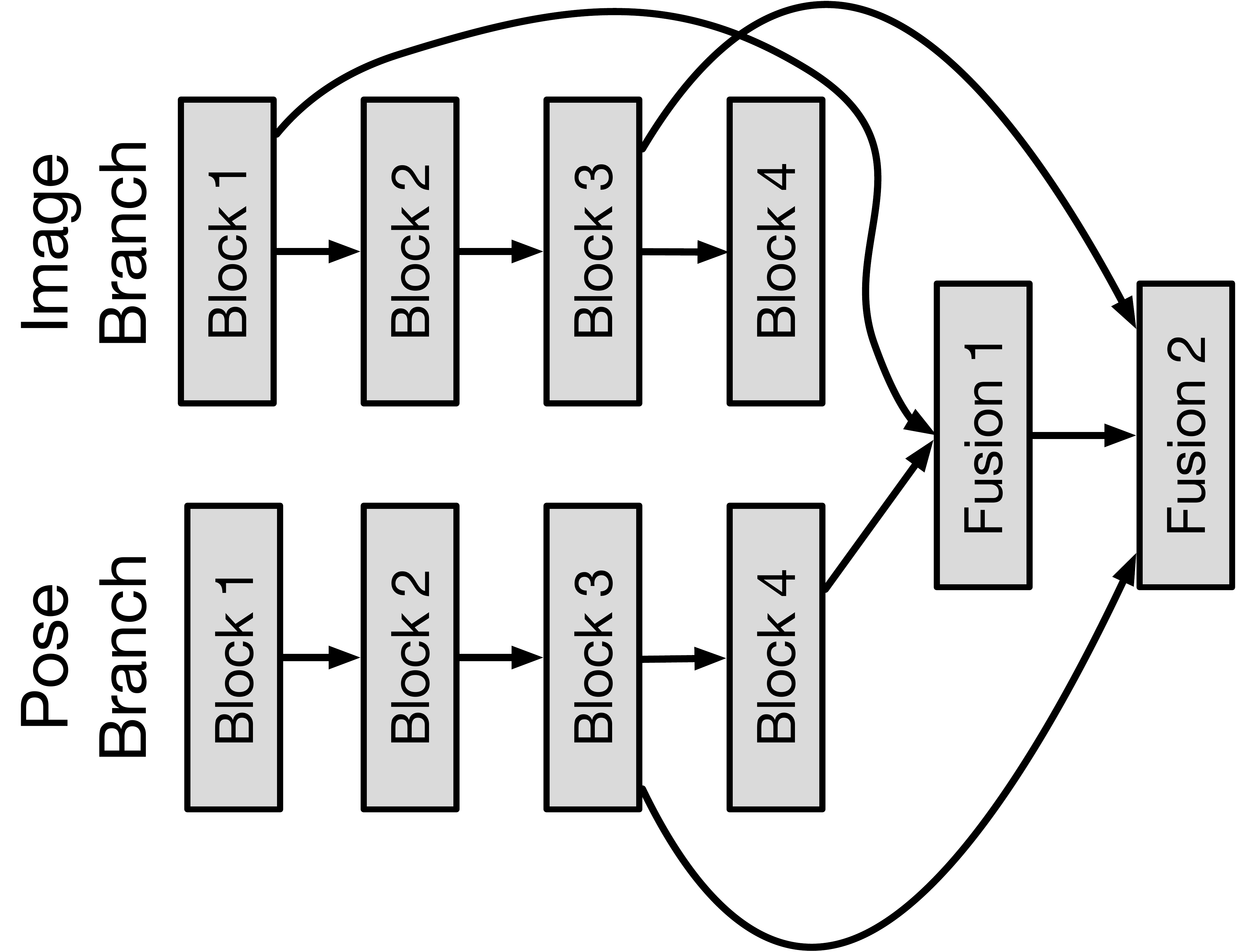}
} 
\caption{Comparison of networks with highest RA generated by NAS using different RA estimation methods.}
\label{fig:nasnet}
\vspace{-10pt}
\end{figure}
}

\no{\paragraph{SA-vs-RA Comparison.}
Comparing the SA and RA of the top networks predicted by the surrogate function as $L$ increases from 1 to 3 (the SA and RA are actual values rather than predicted), we observe a linear correlation between the two metrics, indicating that improving one improves the other too, a useful empirical observation when designing fusion DNNs. That said, there are three networks that stray away from the linear trend, indicating that SA and RA are not always correlated. We omit the detailed data here due to space limit.}


\paragraph{Improving NAS Speed.} Our importance sampling strategy significantly improves the NAS speed by requiring fewer samples in each NAS iteration, which must estimate the RA of the top $K$ network that the surrogate function predicts.

\Fig{fig:latcomp} shows the total NAS time on two Nvidia 2080Ti GPUs across different schemes. ``Baseline'' denotes the NAS time when SA is the search target, i.e., no time spent on RA estimation. The other curves represent the NAS time under different sampling strategies. The $x$-axis shows the number of the fusion layers allowed to be searched. Naturally, when more fusion layers are allowed the total NAS time increases.
Compared to the baseline, \mode{Uniform} sampling almost double the NAS time. \mode{IS-B} limits the overhead to be only 25.5\%.



%% file: relate.tex
\section{Related Work}
\label{sec:related}

The main novelty of our work is two-fold. First, we formulate the new RA metric, which quantifies the accuracy of a network \textit{given} that a fault has occurred. In contrast, FIdelity~\cite{he2020fidelity} estimates the FIT rate; TensorFI~\cite{chen2020tensorfi} and PyTorchFI~\cite{mahmoud2020pytorchfi} both report the SDC rate.
Both the FIT rate and the SDC rate quantify \textit{how often} hardware faults show up in software, rather than how bad the network inference would be under faults.
Mahmoud et al.~\cite{mahmoud2021optimizing} estimates the ``relative vulnerability of each fmap in the CNN'' (for selectively protecting feature maps~\cite{libano2018selective}), a metric different from our RA.

Second, we provide an importance sampling-based method to estimate the RA metric without exhausting all the software fault sites.
Prior work mostly sample faults sites uniformly. Uniform sampling in software  (PyTorchFI~\cite{mahmoud2020pytorchfi} and TensorFI~\cite{chen2020tensorfi}) under-estimates RA (\Sect{sec:compare}).
Uniformly sampling in hardware, e.g., Li et al.~\cite{li2017understanding}, in theory could yield the same RA estimation as ours. But hardware simulation is known to be extremely time consuming, which our work avoids.
Mahmoud et al.~\cite{mahmoud2021optimizing} breaks away from uniform sampling. They instead sample based on the number of MAC operations to calculate each activation. As shown in \Fig{fig:convbignet}, such a method, while significantly better than uniform sampling, converges more slowly than importance sampling.

We use importance sampling with PDF heuristics derived from DNN-specific characteristics. The PDF heuristics are reminiscent of the classic Monte Carlo methods used in solving the rendering equation for physically-based rendering~\cite{pharr2016physically}.





Much of the prior work studies the DNN resiliency against faults in memory~\cite{reagen2018ares, mittal2020survey, chandramoorthy2019resilient}. Memory faults could also be introduced by an attacker~\cite{chen2021proflip,rakin2020tbt,hong2019terminal}. This work focuses on the transient faults in FFs as discussed in \Sect{sec:motivation:metric}. Hong et al.~\cite{hong2019terminal} considers faults just in the most significant bits of a floating point representation and, thus, over-estimates the impact of transient faults.

Transient faults are most accurately studied in hardware~\cite{cheng2016clear, cho2013quantitative,wang2004characterizing}, which is impractical due to slow RTL simulations.
Modeling transient errors at the model level, instead, has been the dominant approach for studying DNN resiliency~\cite{mahmoud2020pytorchfi,chen2020tensorfi,hong2019terminal,chen2019binfi,chen2008error}. We discuss the sources of inaccuracies of this approach in \Sect{sec:motivation:errors}.
FIdelity~\cite{he2020fidelity} provides a methodology that captures the impact of a single hardware fault without RTL simulation, which this paper uses.

Of course, transient faults are not the only source of vulnerability in real-world applications such as autonomous machines, where safety issues arise from latency variation~\cite{zhao2020safety}, adversarial attacks to DNNs~\cite{gan2020ptolemy, qiu2019adversarial}, and others; different sources of vulnerability have different implications on the safety~\cite{gan2022braum}. An interesting line of future work is to examine DNN soft errors in the context of end-to-end applications.

Our RA estimation framework assumes that a DNN's data reuse is regular: all variables of the same type in a layer have the same reuse.
This assumption does not hold for Graph Neural Networks~\cite{scarselli2008graph,shchur2018pitfalls,zhang2019heterogeneous} and point cloud DNNs~\cite{zhang2021point,tiator2020point,zhou2018voxelnet,qi2017pointnet}, where data reuses are not uniform because of the irregular connections in graphs and/or neighbor searches~\cite{tao2002continuous,jegou2010product,seidl1998optimal, xu2019tigris, feng2022crescent, zhu2022rtnn}.

%% file: concl.tex
\section{Conclusion}

We propose a statistically-based formulation to quantify a DNN's RA. The crux of the formulation is to correctly model the faulty probabilities of software fault sites. Our RA formulation can be seen as the DNN-specific AVF.
We show that importance sampling coupled with proper sampling functions is key to efficiently estimating the RA in software.
We show that transient faults present far greater degradation to DNN's inference accuracy than what existing metrics estimate.

%% file: main.bbl
\begin{thebibliography}{10}
\providecommand{\url}[1]{#1}
\csname url@samestyle\endcsname
\providecommand{\newblock}{\relax}
\providecommand{\bibinfo}[2]{#2}
\providecommand{\BIBentrySTDinterwordspacing}{\spaceskip=0pt\relax}
\providecommand{\BIBentryALTinterwordstretchfactor}{4}
\providecommand{\BIBentryALTinterwordspacing}{\spaceskip=\fontdimen2\font plus
\BIBentryALTinterwordstretchfactor\fontdimen3\font minus
  \fontdimen4\font\relax}
\providecommand{\BIBforeignlanguage}[2]{{%
\expandafter\ifx\csname l@#1\endcsname\relax
\typeout{** WARNING: IEEEtranS.bst: No hyphenation pattern has been}%
\typeout{** loaded for the language `#1'. Using the pattern for}%
\typeout{** the default language instead.}%
\else
\language=\csname l@#1\endcsname
\fi
#2}}
\providecommand{\BIBdecl}{\relax}
\BIBdecl

\bibitem{ap}
``{Baidu Apollo team (2017), Apollo: Open Source Autonomous Driving},
  howpublished = {\url{https://github.com/apolloauto/apollo}} note = {Accessed:
  2019-02-11}.''

\bibitem{ev}
``{Expected Value}, howpublished =
  {\url{https://en.wikipedia.org/wiki/expected_value}}.''

\bibitem{nvdla}
``{Nvdla open source project}, howpublished = {\url{
  http://nvdla.org/primer.html}}, note = {Accessed: 2018}.''

\bibitem{z01x}
``{Z01X Functional Safety Assurance}, howpublished =
  {\url{https://www.synopsys.com/verification/simulation/z01x-functional-safety.html}}.''

\bibitem{baradel2018glimpse}
F.~Baradel, C.~Wolf, J.~Mille, and G.~W. Taylor, ``Glimpse clouds: Human
  activity recognition from unstructured feature points,'' in \emph{Proceedings
  of the IEEE Conference on Computer Vision and Pattern Recognition}, 2018, pp.
  469--478.

\bibitem{calin1996upset}
T.~Calin, M.~Nicolaidis, and R.~Velazco, ``Upset hardened memory design for
  submicron cmos technology,'' \emph{IEEE Transactions on nuclear science},
  vol.~43, no.~6, pp. 2874--2878, 1996.

\bibitem{carion2020end}
N.~Carion, F.~Massa, G.~Synnaeve, N.~Usunier, A.~Kirillov, and S.~Zagoruyko,
  ``End-to-end object detection with transformers,'' in \emph{European
  conference on computer vision}.\hskip 1em plus 0.5em minus 0.4em\relax
  Springer, 2020, pp. 213--229.

\bibitem{chandramoorthy2019resilient}
N.~Chandramoorthy, K.~Swaminathan, M.~Cochet, A.~Paidimarri, S.~Eldridge, R.~V.
  Joshi, M.~M. Ziegler, A.~Buyuktosunoglu, and P.~Bose, ``Resilient low voltage
  accelerators for high energy efficiency,'' in \emph{2019 IEEE International
  Symposium on High Performance Computer Architecture (HPCA)}.\hskip 1em plus
  0.5em minus 0.4em\relax IEEE, 2019, pp. 147--158.

\bibitem{chen1984error}
C.-L. Chen and M.~Hsiao, ``Error-correcting codes for semiconductor memory
  applications: A state-of-the-art review,'' \emph{IBM Journal of Research and
  development}, vol.~28, no.~2, pp. 124--134, 1984.

\bibitem{chen2008error}
D.~Chen, G.~Jacques-Silva, Z.~Kalbarczyk, R.~K. Iyer, and B.~Mealey, ``Error
  behavior comparison of multiple computing systems: A case study using linux
  on pentium, solaris on sparc, and aix on power,'' in \emph{2008 14th IEEE
  Pacific Rim International Symposium on Dependable Computing}.\hskip 1em plus
  0.5em minus 0.4em\relax IEEE, 2008, pp. 339--346.

\bibitem{chen2021proflip}
H.~Chen, C.~Fu, J.~Zhao, and F.~Koushanfar, ``Proflip: Targeted trojan attack
  with progressive bit flips,'' in \emph{Proceedings of the IEEE/CVF
  International Conference on Computer Vision}, 2021, pp. 7718--7727.

\bibitem{chen2019binfi}
Z.~Chen, G.~Li, K.~Pattabiraman, and N.~DeBardeleben, ``Binfi: An efficient
  fault injector for safety-critical machine learning systems,'' in
  \emph{Proceedings of the International Conference for High Performance
  Computing, Networking, Storage and Analysis}, 2019, pp. 1--23.

\bibitem{chen2020tensorfi}
Z.~Chen, N.~Narayanan, B.~Fang, G.~Li, K.~Pattabiraman, and N.~DeBardeleben,
  ``Tensorfi: A flexible fault injection framework for tensorflow
  applications,'' in \emph{2020 IEEE 31st International Symposium on Software
  Reliability Engineering (ISSRE)}.\hskip 1em plus 0.5em minus 0.4em\relax
  IEEE, 2020, pp. 426--435.

\bibitem{cheng2016clear}
E.~Cheng, S.~Mirkhani, L.~G. Szafaryn, C.-Y. Cher, H.~Cho, K.~Skadron, M.~R.
  Stan, K.~Lilja, J.~A. Abraham, P.~Bose \emph{et~al.}, ``Clear: C ross-l ayer
  e xploration for a rchitecting r esilience-combining hardware and software
  techniques to tolerate soft errors in processor cores,'' in \emph{Proceedings
  of the 53rd Annual Design Automation Conference}, 2016, pp. 1--6.

\bibitem{cho2013quantitative}
H.~Cho, S.~Mirkhani, C.-Y. Cher, J.~A. Abraham, and S.~Mitra, ``Quantitative
  evaluation of soft error injection techniques for robust system design,'' in
  \emph{Proceedings of the 50th Annual Design Automation Conference}, 2013, pp.
  1--10.

\bibitem{de2015evaluation}
D.~A. G.~G. de~Oliveira, L.~L. Pilla, T.~Santini, and P.~Rech, ``Evaluation and
  mitigation of radiation-induced soft errors in graphics processing units,''
  \emph{IEEE Transactions on Computers}, vol.~65, no.~3, pp. 791--804, 2015.

\bibitem{degalahal2003analyzing}
V.~Degalahal, N.~Vijaykrishnan, and M.~J. Irwin, ``Analyzing soft errors in
  leakage optimized sram design,'' in \emph{16th International Conference on
  VLSI Design, 2003. Proceedings.}\hskip 1em plus 0.5em minus 0.4em\relax IEEE,
  2003, pp. 227--233.

\bibitem{deng2009imagenet}
J.~Deng, W.~Dong, R.~Socher, L.-J. Li, K.~Li, and L.~Fei-Fei, ``Imagenet: A
  large-scale hierarchical image database,'' in \emph{2009 IEEE conference on
  computer vision and pattern recognition}.\hskip 1em plus 0.5em minus
  0.4em\relax Ieee, 2009, pp. 248--255.

\bibitem{deng2012mnist}
L.~Deng, ``The mnist database of handwritten digit images for machine learning
  research,'' \emph{IEEE Signal Processing Magazine}, vol.~29, no.~6, pp.
  141--142, 2012.

\bibitem{dos2021revealing}
F.~F. dos Santos, J.~E.~R. Condia, L.~Carro, M.~S. Reorda, and P.~Rech,
  ``Revealing gpus vulnerabilities by combining register-transfer and
  software-level fault injection,'' in \emph{2021 51st Annual IEEE/IFIP
  International Conference on Dependable Systems and Networks (DSN)}.\hskip 1em
  plus 0.5em minus 0.4em\relax IEEE, 2021, pp. 292--304.

\bibitem{dos2018analyzing}
F.~F. dos Santos, P.~F. Pimenta, C.~Lunardi, L.~Draghetti, L.~Carro, D.~Kaeli,
  and P.~Rech, ``Analyzing and increasing the reliability of convolutional
  neural networks on gpus,'' \emph{IEEE Transactions on Reliability}, vol.~68,
  no.~2, pp. 663--677, 2018.

\bibitem{feng2010shoestring}
S.~Feng, S.~Gupta, A.~Ansari, and S.~Mahlke, ``Shoestring: probabilistic soft
  error reliability on the cheap,'' \emph{ACM SIGARCH Computer Architecture
  News}, vol.~38, no.~1, pp. 385--396, 2010.

\bibitem{feng2022crescent}
Y.~Feng, G.~Hammonds, Y.~Gan, and Y.~Zhu, ``Crescent: taming memory
  irregularities for accelerating deep point cloud analytics,'' \emph{arXiv
  preprint arXiv:2204.10707}, 2022.

\bibitem{gan2020ptolemy}
Y.~Gan, Y.~Qiu, J.~Leng, M.~Guo, and Y.~Zhu, ``Ptolemy: Architecture support
  for robust deep learning,'' in \emph{2020 53rd Annual IEEE/ACM International
  Symposium on Microarchitecture (MICRO)}.\hskip 1em plus 0.5em minus
  0.4em\relax IEEE, 2020, pp. 241--255.

\bibitem{gan2022braum}
Y.~Gan, P.~Whatmough, J.~Leng, B.~Yu, S.~Liu, and Y.~Zhu, ``Braum: Analyzing
  and protecting autonomous machine software stack,'' in \emph{ISSRE}, 2022.

\bibitem{hannun2014deep}
A.~Hannun, C.~Case, J.~Casper, B.~Catanzaro, G.~Diamos, E.~Elsen, R.~Prenger,
  S.~Satheesh, S.~Sengupta, A.~Coates \emph{et~al.}, ``Deep speech: Scaling up
  end-to-end speech recognition,'' \emph{arXiv preprint arXiv:1412.5567}, 2014.

\bibitem{hari2012low}
S.~K.~S. Hari, S.~V. Adve, and H.~Naeimi, ``Low-cost program-level detectors
  for reducing silent data corruptions,'' in \emph{IEEE/IFIP international
  conference on dependable systems and networks (DSN 2012)}.\hskip 1em plus
  0.5em minus 0.4em\relax IEEE, 2012, pp. 1--12.

\bibitem{he2016deep}
K.~He, X.~Zhang, S.~Ren, and J.~Sun, ``Deep residual learning for image
  recognition,'' in \emph{Proceedings of the IEEE conference on computer vision
  and pattern recognition}, 2016, pp. 770--778.

\bibitem{he2020fidelity}
Y.~He, P.~Balaprakash, and Y.~Li, ``Fidelity: Efficient resilience analysis
  framework for deep learning accelerators,'' in \emph{2020 53rd Annual
  IEEE/ACM International Symposium on Microarchitecture (MICRO)}.\hskip 1em
  plus 0.5em minus 0.4em\relax IEEE, 2020, pp. 270--281.

\bibitem{hong2019terminal}
S.~Hong, P.~Frigo, Y.~Kaya, C.~Giuffrida, and T.~Dumitraș, ``Terminal brain
  damage: Exposing the graceless degradation in deep neural networks under
  hardware fault attacks,'' in \emph{28th USENIX Security Symposium (USENIX
  Security 19)}, 2019, pp. 497--514.

\bibitem{hosang2015taking}
J.~Hosang, M.~Omran, R.~Benenson, and B.~Schiele, ``Taking a deeper look at
  pedestrians,'' in \emph{Proceedings of the IEEE conference on computer vision
  and pattern recognition}, 2015, pp. 4073--4082.

\bibitem{howard2017mobilenets}
A.~G. Howard, M.~Zhu, B.~Chen, D.~Kalenichenko, W.~Wang, T.~Weyand,
  M.~Andreetto, and H.~Adam, ``Mobilenets: Efficient convolutional neural
  networks for mobile vision applications,'' \emph{arXiv preprint
  arXiv:1704.04861}, 2017.

\bibitem{jagannathan2012frequency}
S.~Jagannathan, T.~Loveless, B.~Bhuva, N.~Gaspard, N.~Mahatme, T.~Assis, S.-J.
  Wen, R.~Wong, and L.~Massengill, ``Frequency dependence of alpha-particle
  induced soft error rates of flip-flops in 40-nm cmos technology,'' \emph{IEEE
  Transactions on Nuclear Science}, vol.~59, no.~6, pp. 2796--2802, 2012.

\bibitem{jahinuzzaman2009soft}
S.~M. Jahinuzzaman, D.~J. Rennie, and M.~Sachdev, ``A soft error tolerant 10t
  sram bit-cell with differential read capability,'' \emph{IEEE Transactions on
  Nuclear Science}, vol.~56, no.~6, pp. 3768--3773, 2009.

\bibitem{jegou2010product}
H.~Jegou, M.~Douze, and C.~Schmid, ``Product quantization for nearest neighbor
  search,'' \emph{IEEE transactions on pattern analysis and machine
  intelligence}, vol.~33, no.~1, pp. 117--128, 2010.

\bibitem{kato2018autoware}
S.~Kato, S.~Tokunaga, Y.~Maruyama, S.~Maeda, M.~Hirabayashi, Y.~Kitsukawa,
  A.~Monrroy, T.~Ando, Y.~Fujii, and T.~Azumi, ``Autoware on board: Enabling
  autonomous vehicles with embedded systems,'' in \emph{Proceedings of the 9th
  ACM/IEEE International Conference on Cyber-Physical Systems (ICCPS)}, 2018,
  pp. 287--296.

\bibitem{kloek1978bayesian}
T.~Kloek and H.~K. Van~Dijk, ``Bayesian estimates of equation system
  parameters: an application of integration by monte carlo,''
  \emph{Econometrica: Journal of the Econometric Society}, pp. 1--19, 1978.

\bibitem{kobayashi2014low}
K.~Kobayashi, K.~Kubota, M.~Masuda, Y.~Manzawa, J.~Furuta, S.~Kanda, and
  H.~Onodera, ``A low-power and area-efficient radiation-hard redundant
  flip-flop, dice acff, in a 65 nm thin-box fd-soi,'' \emph{IEEE Transactions
  on Nuclear Science}, vol.~61, no.~4, pp. 1881--1888, 2014.

\bibitem{krizhevsky2009learning}
A.~Krizhevsky, G.~Hinton \emph{et~al.}, ``Learning multiple layers of features
  from tiny images,'' 2009.

\bibitem{NIPS2012_c399862d}
\BIBentryALTinterwordspacing
A.~Krizhevsky, I.~Sutskever, and G.~E. Hinton, ``Imagenet classification with
  deep convolutional neural networks,'' in \emph{Advances in Neural Information
  Processing Systems}, F.~Pereira, C.~Burges, L.~Bottou, and K.~Weinberger,
  Eds., vol.~25.\hskip 1em plus 0.5em minus 0.4em\relax Curran Associates,
  Inc., 2012. [Online]. Available:
  \url{https://proceedings.neurips.cc/paper/2012/file/c399862d3b9d6b76c8436e924a68c45b-Paper.pdf}
\BIBentrySTDinterwordspacing

\bibitem{kroese2013handbook}
D.~P. Kroese, T.~Taimre, and Z.~I. Botev, \emph{Handbook of monte carlo
  methods}.\hskip 1em plus 0.5em minus 0.4em\relax John Wiley \& Sons, 2013.

\bibitem{lantz1996soft}
L.~Lantz, ``Soft errors induced by alpha particles,'' \emph{IEEE Transactions
  on Reliability}, vol.~45, no.~2, pp. 174--179, 1996.

\bibitem{leveugle2009statistical}
R.~Leveugle, A.~Calvez, P.~Maistri, and P.~Vanhauwaert, ``Statistical fault
  injection: Quantified error and confidence,'' in \emph{2009 Design,
  Automation \& Test in Europe Conference \& Exhibition}.\hskip 1em plus 0.5em
  minus 0.4em\relax IEEE, 2009, pp. 502--506.

\bibitem{li2018co}
C.~Li, Q.~Zhong, D.~Xie, and S.~Pu, ``Co-occurrence feature learning from
  skeleton data for action recognition and detection with hierarchical
  aggregation,'' \emph{arXiv preprint arXiv:1804.06055}, 2018.

\bibitem{li2017understanding}
G.~Li, S.~K.~S. Hari, M.~Sullivan, T.~Tsai, K.~Pattabiraman, J.~Emer, and S.~W.
  Keckler, ``Understanding error propagation in deep learning neural network
  (dnn) accelerators and applications,'' in \emph{Proceedings of the
  International Conference for High Performance Computing, Networking, Storage
  and Analysis}, 2017, pp. 1--12.

\bibitem{libano2018selective}
F.~Libano, B.~Wilson, J.~Anderson, M.~J. Wirthlin, C.~Cazzaniga, C.~Frost, and
  P.~Rech, ``Selective hardening for neural networks in fpgas,'' \emph{IEEE
  Transactions on Nuclear Science}, vol.~66, no.~1, pp. 216--222, 2018.

\bibitem{lin2014microsoft}
T.-Y. Lin, M.~Maire, S.~Belongie, J.~Hays, P.~Perona, D.~Ramanan,
  P.~Doll{\'a}r, and C.~L. Zitnick, ``Microsoft coco: Common objects in
  context,'' in \emph{European conference on computer vision}.\hskip 1em plus
  0.5em minus 0.4em\relax Springer, 2014, pp. 740--755.

\bibitem{lunardi2018efficacy}
C.~Lunardi, F.~Previlon, D.~Kaeli, and P.~Rech, ``On the efficacy of ecc and
  the benefits of finfet transistor layout for gpu reliability,'' \emph{IEEE
  Transactions on Nuclear Science}, vol.~65, no.~8, pp. 1843--1850, 2018.

\bibitem{madeira2000emulation}
H.~Madeira, D.~Costa, and M.~Vieira, ``On the emulation of software faults by
  software fault injection,'' in \emph{Proceeding International Conference on
  Dependable Systems and Networks. DSN 2000}.\hskip 1em plus 0.5em minus
  0.4em\relax IEEE, 2000, pp. 417--426.

\bibitem{mahmoud2020pytorchfi}
A.~Mahmoud, N.~Aggarwal, A.~Nobbe, J.~R.~S. Vicarte, S.~V. Adve, C.~W.
  Fletcher, I.~Frosio, and S.~K.~S. Hari, ``Pytorchfi: A runtime perturbation
  tool for dnns,'' in \emph{2020 50th Annual IEEE/IFIP International Conference
  on Dependable Systems and Networks Workshops (DSN-W)}.\hskip 1em plus 0.5em
  minus 0.4em\relax IEEE, 2020, pp. 25--31.

\bibitem{mahmoud2021optimizing}
A.~Mahmoud, S.~K.~S. Hari, C.~W. Fletcher, S.~V. Adve, C.~Sakr, N.~Shanbhag,
  P.~Molchanov, M.~B. Sullivan, T.~Tsai, and S.~W. Keckler, ``Optimizing
  selective protection for cnn resilience,'' in \emph{32nd IEEE International
  Symposium on Software Reliability Engineering, ISSRE 2021}.\hskip 1em plus
  0.5em minus 0.4em\relax IEEE Computer Society, 2021, pp. 127--138.

\bibitem{mittal2020survey}
S.~Mittal, ``A survey on modeling and improving reliability of dnn algorithms
  and accelerators,'' \emph{Journal of Systems Architecture}, vol. 104, p.
  101689, 2020.

\bibitem{mukherjee2003systematic}
S.~S. Mukherjee, C.~Weaver, J.~Emer, S.~K. Reinhardt, and T.~Austin, ``A
  systematic methodology to compute the architectural vulnerability factors for
  a high-performance microprocessor,'' in \emph{Proceedings. 36th Annual
  IEEE/ACM International Symposium on Microarchitecture, 2003. MICRO-36.}\hskip
  1em plus 0.5em minus 0.4em\relax IEEE, 2003, pp. 29--40.

\bibitem{naseer2007critical}
R.~Naseer, Y.~Boulghassoul, J.~Draper, S.~DasGupta, and A.~Witulski, ``Critical
  charge characterization for soft error rate modeling in 90nm sram,'' in
  \emph{2007 IEEE International Symposium on Circuits and Systems}.\hskip 1em
  plus 0.5em minus 0.4em\relax IEEE, 2007, pp. 1879--1882.

\bibitem{panayotov2015librispeech}
V.~Panayotov, G.~Chen, D.~Povey, and S.~Khudanpur, ``Librispeech: an asr corpus
  based on public domain audio books,'' in \emph{2015 IEEE international
  conference on acoustics, speech and signal processing (ICASSP)}.\hskip 1em
  plus 0.5em minus 0.4em\relax IEEE, 2015, pp. 5206--5210.

\bibitem{papadimitriou2021demystifying}
G.~Papadimitriou and D.~Gizopoulos, ``Demystifying the system vulnerability
  stack: Transient fault effects across the layers,'' in \emph{2021 ACM/IEEE
  48th Annual International Symposium on Computer Architecture (ISCA)}.\hskip
  1em plus 0.5em minus 0.4em\relax IEEE, 2021, pp. 902--915.

\bibitem{NEURIPS2019_9015}
A.~Paszke, S.~Gross, F.~Massa, A.~Lerer, J.~Bradbury, G.~Chanan, T.~Killeen,
  Z.~Lin, N.~Gimelshein, L.~Antiga \emph{et~al.}, ``Pytorch: An imperative
  style, high-performance deep learning library,'' \emph{Advances in neural
  information processing systems}, vol.~32, 2019.

\bibitem{peng2018radiation}
C.~Peng, J.~Huang, C.~Liu, Q.~Zhao, S.~Xiao, X.~Wu, Z.~Lin, J.~Chen, and
  X.~Zeng, ``Radiation-hardened 14t sram bitcell with speed and power optimized
  for space application,'' \emph{IEEE Transactions on Very Large Scale
  Integration (VLSI) Systems}, vol.~27, no.~2, pp. 407--415, 2018.

\bibitem{perez2019mfas}
J.-M. P{\'e}rez-R{\'u}a, V.~Vielzeuf, S.~Pateux, M.~Baccouche, and F.~Jurie,
  ``Mfas: Multimodal fusion architecture search,'' in \emph{Proceedings of the
  IEEE/CVF Conference on Computer Vision and Pattern Recognition}, 2019, pp.
  6966--6975.

\bibitem{pharr2016physically}
M.~Pharr, W.~Jakob, and G.~Humphreys, \emph{Physically based rendering: From
  theory to implementation}.\hskip 1em plus 0.5em minus 0.4em\relax Morgan
  Kaufmann, 2016.

\bibitem{press2007numerical}
W.~H. Press, S.~A. Teukolsky, W.~T. Vetterling, and B.~P. Flannery,
  \emph{Numerical recipes 3rd edition: The art of scientific computing}.\hskip
  1em plus 0.5em minus 0.4em\relax Cambridge university press, 2007.

\bibitem{qi2017pointnet}
C.~R. Qi, H.~Su, K.~Mo, and L.~J. Guibas, ``Pointnet: Deep learning on point
  sets for 3d classification and segmentation,'' in \emph{Proceedings of the
  IEEE conference on computer vision and pattern recognition}, 2017, pp.
  652--660.

\bibitem{qiu2019adversarial}
Y.~Qiu, J.~Leng, C.~Guo, Q.~Chen, C.~Li, M.~Guo, and Y.~Zhu, ``Adversarial
  defense through network profiling based path extraction,'' in
  \emph{Proceedings of the IEEE/CVF Conference on Computer Vision and Pattern
  Recognition}, 2019, pp. 4777--4786.

\bibitem{rajaei2015single}
R.~Rajaei, M.~Tabandeh, and M.~Fazeli, ``Single event multiple upset (semu)
  tolerant latch designs in presence of process and temperature variations,''
  \emph{Journal of Circuits, Systems and Computers}, vol.~24, no.~01, p.
  1550007, 2015.

\bibitem{rakin2020tbt}
A.~S. Rakin, Z.~He, and D.~Fan, ``Tbt: Targeted neural network attack with bit
  trojan,'' in \emph{Proceedings of the IEEE/CVF Conference on Computer Vision
  and Pattern Recognition}, 2020, pp. 13\,198--13\,207.

\bibitem{reagen2018ares}
B.~Reagen, U.~Gupta, L.~Pentecost, P.~Whatmough, S.~K. Lee, N.~Mulholland,
  D.~Brooks, and G.-Y. Wei, ``Ares: A framework for quantifying the resilience
  of deep neural networks,'' in \emph{2018 55th ACM/ESDA/IEEE Design Automation
  Conference (DAC)}.\hskip 1em plus 0.5em minus 0.4em\relax IEEE, 2018, pp.
  1--6.

\bibitem{rech2022reliability}
R.~L. Rech and P.~Rech, ``Reliability of google's tensor processing units for
  embedded applications,'' in \emph{2022 Design, Automation \& Test in Europe
  Conference \& Exhibition (DATE)}.\hskip 1em plus 0.5em minus 0.4em\relax
  IEEE, 2022, pp. 376--381.

\bibitem{redmon2018yolov3}
J.~Redmon and A.~Farhadi, ``Yolov3: An incremental improvement,'' \emph{arXiv
  preprint arXiv:1804.02767}, 2018.

\bibitem{scarselli2008graph}
F.~Scarselli, M.~Gori, A.~C. Tsoi, M.~Hagenbuchner, and G.~Monfardini, ``The
  graph neural network model,'' \emph{IEEE transactions on neural networks},
  vol.~20, no.~1, pp. 61--80, 2008.

\bibitem{seidl1998optimal}
T.~Seidl and H.-P. Kriegel, ``Optimal multi-step k-nearest neighbor search,''
  in \emph{Proceedings of the 1998 ACM SIGMOD international conference on
  Management of data}, 1998, pp. 154--165.

\bibitem{seifert2010radiation}
N.~Seifert, V.~Ambrose, B.~Gill, Q.~Shi, R.~Allmon, C.~Recchia, S.~Mukherjee,
  N.~Nassif, J.~Krause, J.~Pickholtz \emph{et~al.}, ``On the radiation-induced
  soft error performance of hardened sequential elements in advanced bulk cmos
  technologies,'' in \emph{2010 IEEE International Reliability Physics
  Symposium}.\hskip 1em plus 0.5em minus 0.4em\relax IEEE, 2010, pp. 188--197.

\bibitem{seifert2006radiation}
N.~Seifert, P.~Slankard, M.~Kirsch, B.~Narasimham, V.~Zia, C.~Brookreson,
  A.~Vo, S.~Mitra, B.~Gill, and J.~Maiz, ``Radiation-induced soft error rates
  of advanced cmos bulk devices,'' in \emph{2006 IEEE International Reliability
  Physics Symposium Proceedings}.\hskip 1em plus 0.5em minus 0.4em\relax IEEE,
  2006, pp. 217--225.

\bibitem{shahroudy2016ntu}
A.~Shahroudy, J.~Liu, T.-T. Ng, and G.~Wang, ``Ntu rgb+ d: A large scale
  dataset for 3d human activity analysis,'' in \emph{Proceedings of the IEEE
  conference on computer vision and pattern recognition}, 2016, pp. 1010--1019.

\bibitem{shchur2018pitfalls}
O.~Shchur, M.~Mumme, A.~Bojchevski, and S.~G{\"u}nnemann, ``Pitfalls of graph
  neural network evaluation,'' \emph{arXiv preprint arXiv:1811.05868}, 2018.

\bibitem{slayman2005cache}
C.~W. Slayman, ``Cache and memory error detection, correction, and reduction
  techniques for terrestrial servers and workstations,'' \emph{IEEE
  Transactions on Device and Materials Reliability}, vol.~5, no.~3, pp.
  397--404, 2005.

\bibitem{sridharan2009eliminating}
V.~Sridharan and D.~R. Kaeli, ``Eliminating microarchitectural dependency from
  architectural vulnerability,'' in \emph{2009 IEEE 15th International
  Symposium on High Performance Computer Architecture}.\hskip 1em plus 0.5em
  minus 0.4em\relax IEEE, 2009, pp. 117--128.

\bibitem{srinivasan1994accurate}
G.~Srinivasan, P.~Murley, and H.~Tang, ``Accurate, predictive modeling of soft
  error rate due to cosmic rays and chip alpha radiation,'' in
  \emph{Proceedings of 1994 IEEE International Reliability Physics
  Symposium}.\hskip 1em plus 0.5em minus 0.4em\relax IEEE, 1994, pp. 12--16.

\bibitem{tao2002continuous}
Y.~Tao, D.~Papadias, and Q.~Shen, ``Continuous nearest neighbor search,'' in
  \emph{VLDB'02: Proceedings of the 28th International Conference on Very Large
  Databases}.\hskip 1em plus 0.5em minus 0.4em\relax Elsevier, 2002, pp.
  287--298.

\bibitem{tiator2020point}
M.~Tiator, C.~Geiger, and P.~Grimm, ``Point cloud segmentation with deep
  reinforcement learning,'' in \emph{ECAI 2020}.\hskip 1em plus 0.5em minus
  0.4em\relax IOS Press, 2020, pp. 2768--2775.

\bibitem{wang2004characterizing}
N.~J. Wang, J.~Quek, T.~M. Rafacz \emph{et~al.}, ``Characterizing the effects
  of transient faults on a high-performance processor pipeline,'' in
  \emph{International Conference on Dependable Systems and Networks,
  2004}.\hskip 1em plus 0.5em minus 0.4em\relax IEEE Computer Society, 2004,
  pp. 61--61.

\bibitem{wei2014quantifying}
J.~Wei, A.~Thomas, G.~Li, and K.~Pattabiraman, ``Quantifying the accuracy of
  high-level fault injection techniques for hardware faults,'' in \emph{2014
  44th Annual IEEE/IFIP International Conference on Dependable Systems and
  Networks}.\hskip 1em plus 0.5em minus 0.4em\relax IEEE, 2014, pp. 375--382.

\bibitem{xu2019tigris}
T.~Xu, B.~Tian, and Y.~Zhu, ``Tigris: Architecture and algorithms for 3d
  perception in point clouds,'' in \emph{Proceedings of the 52nd Annual
  IEEE/ACM International Symposium on Microarchitecture}, 2019, pp. 629--642.

\bibitem{zhang2019heterogeneous}
C.~Zhang, D.~Song, C.~Huang, A.~Swami, and N.~V. Chawla, ``Heterogeneous graph
  neural network,'' in \emph{Proceedings of the 25th ACM SIGKDD International
  Conference on Knowledge Discovery \& Data Mining}, 2019, pp. 793--803.

\bibitem{zhang2021point}
J.-F. Zhang and Z.~Zhang, ``Point-x: A spatial-locality-aware architecture for
  energy-efficient graph-based point-cloud deep learning,'' in \emph{MICRO-54:
  54th Annual IEEE/ACM International Symposium on Microarchitecture}, 2021, pp.
  1078--1090.

\bibitem{zhao2020safety}
H.~Zhao, Y.~Zhang, P.~Meng, H.~Shi, L.~E. Li, T.~Lou, and J.~Zhao, ``Safety
  score: A quantitative approach to guiding safety-aware autonomous vehicle
  computing system design,'' in \emph{2020 IEEE Intelligent Vehicles Symposium
  (IV)}.\hskip 1em plus 0.5em minus 0.4em\relax IEEE, 2020, pp. 1479--1485.

\bibitem{zhou2018voxelnet}
Y.~Zhou and O.~Tuzel, ``Voxelnet: End-to-end learning for point cloud based 3d
  object detection,'' in \emph{Proceedings of the IEEE conference on computer
  vision and pattern recognition}, 2018, pp. 4490--4499.

\bibitem{zhu2022rtnn}
Y.~Zhu, ``Rtnn: accelerating neighbor search using hardware ray tracing,'' in
  \emph{Proceedings of the 27th ACM SIGPLAN Symposium on Principles and
  Practice of Parallel Programming}, 2022, pp. 76--89.

\end{thebibliography}
